\documentclass[preprint]{elsarticle}

\usepackage[utf8]{inputenc}
\usepackage{todonotes}
\usepackage{url}
\usepackage{amsmath}
\usepackage{amssymb}
\usepackage{xspace}
\usepackage{hyperref}
\usepackage[capitalize]{cleveref}
\usepackage{siunitx}
\usepackage{pdflscape}
\usepackage{chemformula}
\usepackage{multirow}
\usepackage{subcaption}
\usepackage[section]{placeins}
\usepackage{lineno}
\usepackage{tikz}
\usepackage{float}
\usetikzlibrary{shapes.geometric, arrows}

\definecolor{myred}{RGB}{255,51,76}
\definecolor{myblue}{RGB}{36,79,135}
\tikzstyle{redcircle} = [circle, minimum width=1cm, minimum height=1cm, text centered, draw=white, align=center, fill=myred]
\tikzstyle{greencircle} = [circle, minimum width=1cm, minimum height=1cm, text centered, draw=white, text=white, align=center, fill=myblue]
\tikzstyle{redrectangle} = [rectangle, minimum width=1cm, minimum height=1cm, text centered, draw=white, align=center, fill=myred]
\tikzstyle{greenrectangle} = [rectangle, minimum width=1cm, minimum height=1cm, text centered, draw=white, text=white, align=center, fill=myblue]
\tikzstyle{arrow} = [thick,->,>=stealth]
\tikzstyle{doublearrow} = [thick,<->,>=stealth]

\numberwithin{equation}{section}

\graphicspath{{./}}

\newcommand{\dune}{\textsc{Dune}\xspace}
\newcommand{\dumux}{DuMu\textsuperscript{x}\xspace}
\newcommand{\dunemodule}[1]{\texttt{#1}\xspace}

\newcommand{\n}{\mathbf{n}}
\newcommand{\vel}{\mathbf{v}}

\newcommand{\velthreed}{\mathbf{v}_\mathrm{3D}}

\renewcommand{\div}{\nabla\cdot\!}
\newcommand{\grad}{\nabla\!}

\journal{Journal of Computational Physics}

\begin{document}

\title{Model reduction for coupled free flow over porous media: a hybrid dimensional pore network model approach}
\author[1]{Kilian Weishaupt\corref{cor1}}
\ead{kilian.weishaupt@iws.uni-stuttgart.de}
\author[2]{Alexandros Terzis}
\author[4]{Ioannis Zarikos}
\author[3]{Guang Yang}
\author[4]{Matthijs de Winter}
\author[1]{Rainer Helmig}

\cortext[cor1]{Corresponding author}
\address[1]{Department of Hydromechanics and Modelling of Hydrosystems, University of Stuttgart, Stuttgart, Germany}
\address[2]{Department of Mechanical Engineering, Stanford University, Stanford, California, USA}
\address[3]{School of Mechanical Engineering, Shanghai Jiao Tong University, Shanghai, China}
\address[4]{Environmental Hydrogeology Group, Department of Earth Sciences, Utrecht University, Utrecht, Netherlands}

\begin{abstract}
Modeling coupled systems of free flow adjacent to a porous medium by means of fully resolved Navier-Stokes equations is limited by the immense computational
cost and is thus only feasible for relatively small domains. Model reduction allows to decrease a model's complexity while maintaining an acceptable degree
of accuracy. Starting from a fully resolved three-dimensional numerical model, which is compared to high-resolution micro-PIV experimental data obtained from a previous study
\citep{terzis2019a}, we perform a two-fold model reduction: first, a quasi-3D model incorporating a wall friction term is
successfully compared to the fully resolved model. Second, we employ a pore-network model to account for the porous part of the domain and couple it to the quasi-3D model which still
resolves the free-flow part of the domain \citep{weishaupt2019a}. We have extended this coupling approach here to include slip velocities
at the pore throats intersecting with the free-flow domain. The proposed method is simple, accurate and comes at no additional run-time penalty.
The coupled model deviates by less than \SI{10}{\percent} from the other two model concepts. Several hours of run-time was required
for the three-dimensional model compared to eleven and five minutes necessary for the quasi-3D and coupled model which highlights the benefits of model reduction.
\end{abstract}


\maketitle

\section{Introduction}
\label{sec:intro}

Coupled systems of free flow over a porous medium play an important role in many environmental, biological and technical processes. Examples include
evaporation from soil governed by atmospheric air flow \cite{vanderborght2017a}, intervascular exchange in living tissue \cite{chauhan2011a}, preservation of food \cite{verboven2006a},
fuel cell water management \cite{gurau2009a} or heat exchange systems \cite{yang2018a}. Considerable effort has been spent on modeling these kinds of systems where a discrete resolution of the complex porous geometry such as in Direct Numerical Simulation (DNS) is often not computationally feasible for larger systems. The porous medium can instead be treated in an averaged sense, based on the concept of an REV
\cite{whitaker1999a}.
Following the so-called one-domain approach, one set of equations is used to describe both the free flow and the porous medium \cite{brinkman1949a}.
For the two-domain approach, a domain decomposition is performed where the free flow is usually described by the Navier-Stokes equations while the porous medium is accounted for by a lower-order
model, such as Darcy's law \cite{ochoa1995a, layton2002a, jamet2009a, mosthaf2011a}. Appropriate coupling conditions between the two domains have to be formulated to ensure thermodynamic
consistency \cite{hassanizadeh1989a}. While being computationally efficient, these upscaled models may provide an insufficient degree of detail on the pore scale crucial for
certain applications, e.g., when local saturation patterns at the interface of a drying soil globally affect the system \cite{shahraeeni2012a}. For these situations, a new class of so-called
hybrid dimensional models have been developed \cite{scheibe2015a} which combine the high spatial resolution of pore-scale approaches, such as pore-network models, with the computational
efficiency of REV-scale models. Pore-network models simplify the complex void geometry of the porous medium to a collection of equivalent pore elements and provide a comparatively high degree
of pore-scale accuracy at low computational demand \cite{oostrom2016a}.
Pore-network models have been coupled both to Darcy-type \cite{balhoff2007a, arbogast2007a, mehmani2014a} and free flow models \cite{beyhaghi2016a}. In our previous work \cite{weishaupt2019a},
we have presented a fully monolithic, fully implicit coupled model employing the Navier-Stokes equations in the free-flow region and a pore-network model in the porous domain.
The model was verified against a numerical reference solution for stationary single-phase flow and an example of transient compositional flow over a random network was given.

In previous work \cite{terzis2019a}, we have performed high-resolution micro-Particle Image Velocimetry (micro-PIV) experiments on a polydimethylsiloxane (PDMS) micromodel
comprising a free-flow channel over a regular porous structure at low Reynolds numbers. In this paper, we extend the hybrid-dimensional coupled model, again using a numerical reference
solution for verification. The latter is generated for the same geometry as used in the experiment and compared to the experimental results. We use OpenFoam to fully resolve
the microfluidic model in a three-dimensional manner considering stationary, laminar creeping flow conditions. Following this, a two-fold model reduction
approach is followed: First, the 3D numerical model is simplified to a two-dimensional model where an additional wall friction term is introduced
to mimic three-dimensional flow characteristics. Second, the porous part of the model domain is replaced by a pore-network model. We introduce a novel coupling condition to explicitly account for the slip velocities at the pores intersecting with the model interface.

\section{Experimental Setup}
\label{sec:experiment}
\begin{figure}[H]
\centering
\includegraphics[width=1.0\textwidth]{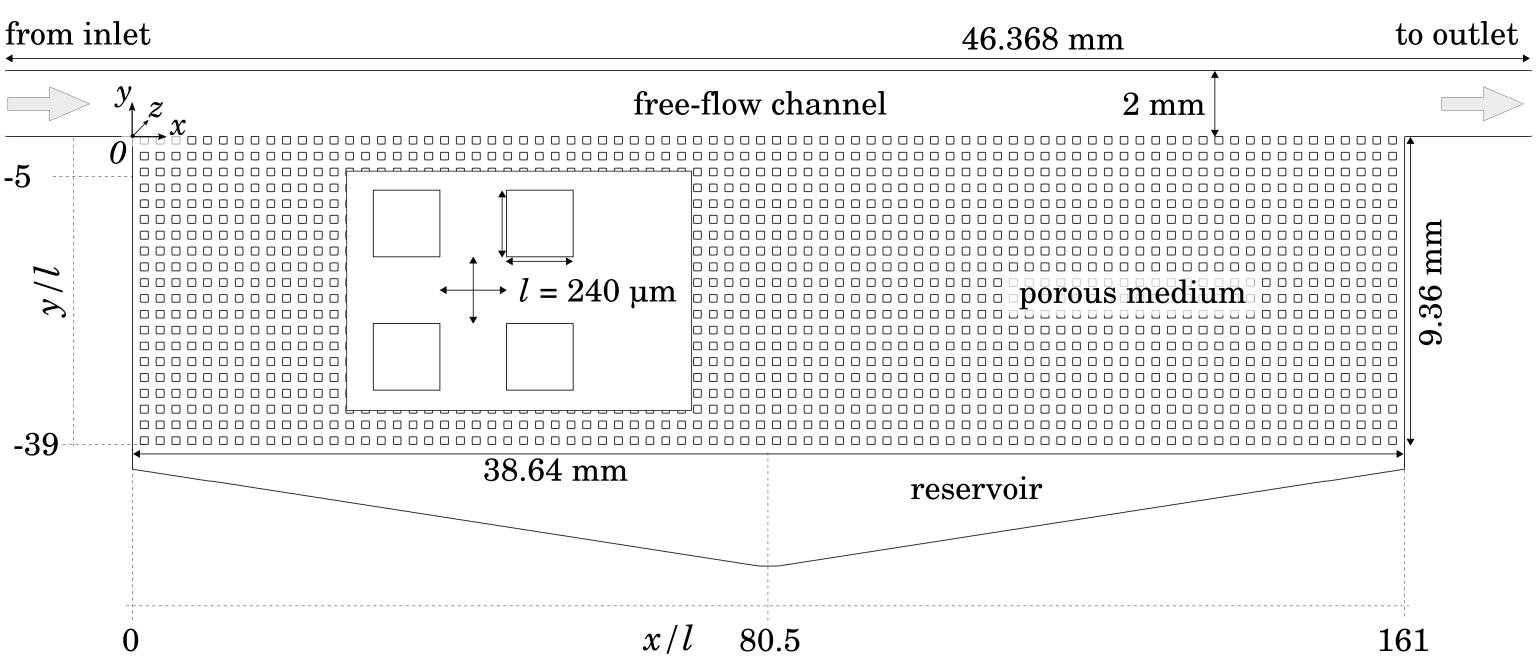}
\caption{Schematic of the PDMS micomodel used in the micro-PIV experiments (redrawn from \cite{terzis2019a}) with dimensions, origin of coordinates and flow direction. The model has a height in %
     $z$-direction of $\SI{200}{\micro \meter}$, the pillars are quadratic with $l = \SI{240}{\micro \meter}$ and evenly spaced throughout the porous domain.}
\label{fig:scheme}
\end{figure}

In this section, we briefly describe the micromodel geometry which will serve as the computational domain for the numerical models.
For details concerning the experimental setup and procedure, we refer to our previous study \citep{terzis2019a}. Fig. \ref{fig:scheme} shows the layout of the PDMS micromodel
which was used in the experiments.
It features three main regions: (1) the free-flow channel at the top, (2) the porous medium  made of 80 x 20 evenly spaced quadratic pillars and (3) a triangular reservoir region
which was included into the design to facilitate the complete saturation of the model with water through an auxiliary inlet (not shown) at the bottom. This inlet was closed during the
experiments. For convenience, two dimensionless lengths $x/l$ and $y/l$ are introduced, where $l = \SI{240e-6}{m}$ which is the width of the pores in the porous region.
The model has a uniform height of $\SI{200e-6}{m}$ in $z$-direction. Note that the inlet and outlet parts of the actual micromodel are longer to ensure a fully developed flow profile
at the beginning of the porous medium. For the simulations, these parts of the channel have been
shortened (and correspond to the dimensions given in the drawing) for efficiency reasons while a fully developed flow was still achieved.

Working under fully saturated conditions, water doped with fluorescent particles was injected at the inlet on the left side of the model ($Re < 1$). Micro-PIV
was used to obtain pore-scale velocity distributions in the porous medium and at the interface region between the free-flow channel and the porous medium.
As the camera's field of view was restricted to 5 x 3 pillars in $x$- and $y$-direction, a series of measurements conducted under steady-state conditions at different locations
of the micromodel was assembled for a full coverage of the model.

\section{Model Concepts}
\label{sec:model}
In this section, we briefly describe the different mathematical numerical models used in this work. Based on the experimental setup, we assume
creeping ($Re < 1$), stationary flow of an incompressible, isothermal fluid and neglect the influence of gravity. The concepts presented can be extended for transient setups and
moderately higher Reynolds numbers (laminar flow) as shown in \cite{weishaupt2019a}. We assume a liquid density of
$\varrho = \SI{1e3}{kg/m^3}$ and a dynamic viscosity of $\mu = \SI{1e-3}{\pascal \second}$.

\subsection{Three-dimensional model (reference model)}
The microfluidic experiments are firstly recalculated in detail on a three-dimensional computational domain that fully resolves the free-flow channel,
the liquid reservoir at the very bottom of the micromodel and all pore-scale structures within the porous domain. We consider the Stokes
equations for incompressible steady-state laminar flow:

\begin{equation}
\label{eq:Stokes}
    \div{[\mu (\grad{ \velthreed } + \grad{ \velthreed }^T)]} -\nabla p = 0 ~.
\end{equation}

The continuity equation is used to close the system:

\begin{equation}
\label{eq:mass-balance}
    \div{\velthreed} = 0 ~.
\end{equation}

The velocity is given as $\velthreed = (v_x, v_y, v_z)^T$.
In the following sections, integrated volume fluxes will be compared. Therefore, we define the volumetric flow in $\si{m^3/s}$ over an area $A$
as:

\begin{equation}
	\label{eq:3d_flow_rate}
    Q_\mathrm{3D} = \int_A (\velthreed  \cdot \mathbf{n}) \mathrm{d} A = \bar{\velthreed } A ~,
\end{equation}

where $\mathbf{n}$ is the unit vector normal to $A$.\\

The open-source CFD tool OpenFoam \citep{openfoam} was used to discretize and solve Eq. \eqref{eq:Stokes} and
\eqref{eq:mass-balance} on a collocated grid using a finite volume approach. Here, the pressure and velocity unknowns both live on the cell centers.

\subsection{Two-dimensional model (quasi-3D model)}
The first reduction step reduces the volume of the model to a two-dimensional plane by dropping the $z$-coordinate as effectively all velocities in $z$-direction are zero for the given geometry.
Under the assumption of a parabolic velocity profile along the $z$-axis of the micromodel,
Eq. \eqref{eq:Stokes} is extended for an additional
drag term  which accounts for the shear forces exerted by the top and bottom wall of the model. This term was first introduced by \cite{flekky1995a} and has been used
successfully for simulating Hele-Shaw flow considering only two dimensions, e.g., in \citep{venturoli2006a, laleian2015a, kunz2015a}:

\begin{equation}
\label{eq:flekkoy}
    \mathbf{f}_\mathrm{drag} = -(8 \mu / h^2) \mathbf{\mathbf{v}_\mathrm{2D}} ~.
\end{equation}

Here, $h$ is the virtual height of flow domain, corresponding to the omitted $z$-coordinate. Ideally, $h \ll w$, where $w$ is the width of a flow
channel, perpendicular to the main flow direction. This leads to Eq. \eqref{eq:Stokes2d}, which now yields the maximum velocity $\mathbf{v}_\mathrm{2D} = (v_x, v_y)^T$
defined on the center plane in $z$-direction:
\begin{equation}
\label{eq:Stokes2d}
        \div{[\mu (\grad{ \mathbf{\mathbf{v}_\mathrm{2D}}} + \grad{ \mathbf{\mathbf{v}_\mathrm{2D}}}^T)]} -\nabla p + \mathbf{f}_\mathrm{drag} = 0 ~,
\end{equation}

Again, the system of equations needs to be closed by a continuity equation:

\begin{equation}
  \label{eq:mass-balance2d}
  \div \mathbf{\mathbf{v}_\mathrm{2D}} = 0 ~.
\end{equation}

In the following, we will refer to this model as \textit{quasi-3D model}. \\
The depth-averaged (over $h$) velocity $\bar{\mathbf{v}}_\mathrm{2D}$ is given by

\begin{equation}
    \bar{\mathbf{v}}_\mathrm{2D} = \frac{2}{3} \mathbf{\mathbf{v}_\mathrm{2D}} ~,
\end{equation}

which then leads to the approximation of the volumetric flow $Q$ in $\si{m^3/s}$ along a given line $s$ which is extruded in $z$-direction by the model's
height $h$:

\begin{equation}
    \label{eq:2d_flow_rate}
    Q = \frac{2}{3} h \int_s (\mathbf{\mathbf{v}_\mathrm{2D}} \cdot \mathbf{n}) \mathrm{d} s ~.
\end{equation}

Here, $\mathbf{n}$ is the unit vector normal to $s$.\\

We use the open-source simulation toolbox \dumux \citep{flemisch2011a, dumux} which is based on the numerical framework \dune \citep{bastian2008a, bastian2008b}
in order to spatially discretize Eq.\eqref{eq:Stokes2d} and \eqref{eq:mass-balance2d}. Here, a finite volume-based staggered grid approach \citep{harlow1965a} is
employed which prevents spurious pressure-velocity decoupling without the need for additional stabilization \citep{versteeg2007a}. This implies that the pressure
degrees of freedom are located on the cell centers while the velocity components live on the cell faces. \dumux is chosen because it provides extensive coupling
capabilities which are required for the hybrid-dimensional model described in the next section. Besides the \dune core modules, \dunemodule{dune-subgrid}
was used in order to create the grid. As the implemented free-flow model also supports Navier-Stokes flow (which is not considered in this work), Newton's method
is chosen by default to solve the potentially non-linear system of equations. \texttt{UMFPack} \citep{davis2004a} is used as direct linear solver.

\subsection{Hybdrid-dimensional model (coupled free flow / pore-network model)}
The final step of model reduction comprises the use of a pore-network model as a proxy for the porous structure. We will continue to use the quasi-3D model in
the free-flow channel above the porous medium and in the triangular reservoir below the latter. For an in-depth description and analysis of the coupled model
we refer to \cite{weishaupt2019a}. Here, we extend this model by considering the drag term given by Eq. \eqref{eq:flekkoy} in the free flow domain and adding
a slip velocity to the pore throats at the interface. For sake of completeness, we briefly discuss the main features of the coupled model followed by a description of the aforementioned model extensions. \\

The free-flow channel and the triangular region are accounted for by using the quasi-3D stokes model described
Eq. \eqref{eq:Stokes2d} and \eqref{eq:mass-balance2d}. \\

In the porous domain, a pore-network model is used where at each pore body (the intersection of two or more pore throats), the continuity
of mass is required:

\begin{equation}
\label{eq:pnm}
 \sum_j Q_{ij} = 0 ~.
\end{equation}

Here $Q_{is}$ is the discrete flow rate in $\si{m^3/s}$ in a throat connection pore bodies $i$ and $j$:

\begin{equation}
\label{eq:pnm-simple}
 Q_{ij} = k_{ij} (p_i - p_j) ~,
\end{equation}

$p_i$ and $p_j$ are the pressures defined at the centers of the pores bodies.
$k_{ij}$ depends on the pore throat geometry and the fluid properties. For certain geometries,
simple analytical expressions for $k_{ij}$ are available in the literature \citep{patzek2001a}. However, for the given porous structure
where there pore throats and bodies have the same dimensions, no suitable expression could be found, thus \cite{weishaupt2019a}
applied a numerical upscaling technique, including also the flow resistance of the pore bodies. We will continue to use this approach here and
refer to the appendix of \citep{weishaupt2019a} for further details. \\

Appropriate coupling conditions are required to ensure the continuity of mass and momentum at the interface between porous medium and free flow \citep{hassanizadeh1989a, layton2002a}.
Here we formulate the coupling conditions for each discrete intersection of a pore throat with the interface. We neglect the subscripts $\mathrm{3D}$ or $\mathrm{2D}$ as
the coupling conditions are valid for both dimensions.

The continuity of flux across the interface is enforced via

\begin{equation}
   [\mathbf{v} \cdot \mathbf{n}]^{\text{FF}} = -[\mathbf{v} \cdot \mathbf{n}]^{\text{PNM}} ~.
\end{equation}

The superscripts $\text{FF}$ and $\text{PNM}$ refer to the interfacial quantities of the free-flow
domain and the pore-network model, respectively.

Compared to \cite{weishaupt2019a}, we revised our coupling conditions for the mechanical equilibrium, i.e., the conservation of momentum across the interface.
We first recall that Eq. \eqref{eq:pnm-simple}, which yields the discrete flux per pore throat in the pore-network model, can be derived from the stationary one-dimensional Stokes equations
\citep{blunt2017a}. Contrary to Darcy-type models \citep{whitaker1999a, layton2002a}, the pore body pressure of the pore-network model has thus the same physical meaning as the pressure of the
Stokes model employed in the free-flow region. Therefore, we require the pressures at the interface to be equal in order to satisfy the balance of forces perpendicular to the interface:

\begin{equation}
\label{eq:coupling_normal}
 [p]^{\text{FF}}  = [p]^{\text{PNM}} ~.
\end{equation}

At the location of solid grains (no intersecting pore throat), a no-slip condition for the free flow is assumed.
Originally \citep{weishaupt2019a}, the tangential component of the discrete pore-throat velocity was used as a slip
condition for the free flow model at the location of the intersecting throats:

\begin{equation}
 [\mathbf{v} \cdot \mathbf{t}_i]^{\text{FF}} = \begin{cases}
 [ \mathbf{v} ]^{\text{PNM}} \cdot [\mathbf{t}_i]^{\text{FF}} ,~ i \in \{0,...,d-1\} \quad \text{on pore throat}\\
0 \hfill\text{else} ~.
\end{cases}
\end{equation}

\begin{equation}
  \label{eq:v_throat}
 [\mathbf{v} ]^{\text{PNM}} = \frac{Q_{ij}}{A_{ij}} \mathbf{n_{ij}} ~,
\end{equation}

$A_{ij}$ is the throat's cross-sectional area while $\mathbf{n_{ij}}$ is a unit normal vector aligned with the throat's central axis, pointing towards the free
flow. The basis of the interface's tangent plane is given by \mbox{$\mathbf{t}_i,~ i \in \{0,...,d-1\}$}.

The disadvantage of this approach is that pore throats intersecting orthogonally with the interface ($\mathbf{n_{ij}} \perp [\mathbf{t}_i]^{\text{FF}}$) will always
feature a no-slip condition at the intersection since \mbox{$[ \mathbf{v} ]^{\text{PNM}} \cdot [\mathbf{t}_i]^{\text{FF}} = 0$}. Here we propose a simple modified approach to approximate the slip velocity at the pore throat on the interface.

We require the continuity of tangential stress

\begin{equation}
 [(-\mu (\grad{ \mathbf{v}} + \grad{ \mathbf{v}}^T) \cdot \n) \cdot \mathbf{t}_i]^{\text{FF}} =  [(-\mu (\grad{ \mathbf{v}} + \grad{ \mathbf{v}}^T) \cdot \n) \cdot \mathbf{t}_i]^{\text{PNM}} ~.
\end{equation}

Instead of trying to calculate the shear rate $\grad{ \mathbf{v}} + \grad{ \mathbf{v}}^T$ in the one-dimensional
pore throats where only uniform, averaged fluxes along the center-line of the throats are defined, we propose a simple
parametrization

\begin{equation}
 [(-\mu (\grad{ \mathbf{v}} + \grad{ \mathbf{v}}^T) \cdot \n) \cdot \mathbf{t}_i]^{\text{FF}} = \beta_\mathrm{throat} \left( [\mathbf{v}]^{\text{FF}} - [\mathbf{v}]^{\text{PNM}} \right) \cdot  \mathbf{t}_i
\end{equation}

in close analogy to the widely-used Beavers-Joseph interface slip condition for REV-scale models \citep{beavers1967a}. The main difference here is that
the slip coefficient $\beta_\mathrm{throat}$ is now defined locally per pore throat and not an averaged quantity of the entire porous medium's interface.

Our new coupling condition for the tangential component of the free-flow velocity thus reads

\begin{equation}
 [\mathbf{v} \cdot \mathbf{t}_i]^{\text{FF}} = \begin{cases}
\vel_{\mathrm{slip},i} \quad \text{on pore throat} ~,\\
0 \hfill\text{else} ~,
\end{cases}
\end{equation}

with

\begin{equation}
 \vel_{\mathrm{slip},i} = \frac{1}{\beta_\mathrm{throat}} [(-\mu (\grad{ \mathbf{v}} + \grad{ \mathbf{v}}^T) \cdot \n) \cdot \mathbf{t}_i]^{\text{FF}} + [\mathbf{v}]^{\text{PNM}} \cdot  \mathbf{t}_i ~.
\end{equation}

We will numerically determine $\beta_\mathrm{throat}$ for various throat widths, as shown later in Sec. \ref{sec:upscaling}.
This approach may be comparable to the work of \cite{bae2016a} who likewise used numerical data to determine a effective slip parameter, yet for non-creeping flow. We furthermore note that there exist analytical solutions to related problems, such as presented in \cite{moffatt1964a, jeong2001a}, which, however, often base on
a number of assumptions and prerequisites which might not always be met in this work.

The coupled model is implemented in a fully monolithic way in \dumux which means that there is only one single system matrix to be solved and no coupling iterations
between the submodels are required \citep{weishaupt2019a}. In addition to the \dune modules mentioned above, we use \dunemodule{dune-foamgrid} \citep{sander2015a}
for the pore-network model. The original model's restriction to odd numbers of free-flow grid cells assigned to each pore throat has been lifted here.

\section{Results and Discussion}
\label{sec:results}
In this section, the results of the numerical simulations conducted with the different modeling approaches are evaluated and discussed. Furthermore, the numerical
evaluation of the pore-scale slip parameter $\beta_\mathrm{thoat}$ is discussed. In all simulations, a fixed fluid viscosity of $\mu = \SI{1e-3}{\pascal \second}$
and density \mbox{$\varrho = \SI{1e3}{\kg / m^3}$} is used. The presence of creeping flow in the experiments makes it sufficient to perform comparisons based on normalized values as long as
$Re < 1$ is satisfied. This implies that different flow rates $Q_1$ and $Q_2$ applied at the inlet of the model result in the same velocity and pressure distributions scaled linearly by a
constant factor $Q_1/Q_2$. Furthermore, preliminary simulations showed that applying fixed-pressure boundary conditions ($\frac{\partial \vel }{\partial x} = 0$) at the inlet and outlet of the
free flow channel decreases the required start-up length for establishing a fully developed velocity profile, thus saving computational cost.
Since no pressures were measured in the experiments, we assign a somehow arbitrary pressure drop $\Delta p = \SI{1e-3}{\pascal}$ between the inlet and the outlet of the free flow channel
for all numerical models presented here, which always yields $Re \ll 1$ based on both the free-flow channel height and the width of the pore throats.

\subsection{3D simulation results and comparison with micro-PIV experimental data}

The geometry of the micromodel described in \cite{terzis2019a} is meshed discretely using regular, axis-parallel cells in order to create a numerical reference solution against which the quasi-3D model can be compared.
Having assured grid convergence, a mesh resolution is chosen such that each pore throat is discretized with 20 cells in all directions
($\Delta x = \Delta y = \SI{1.2e-5}{\meter}, \Delta z = \SI{1.0e-5}{\meter}$), resulting in a total number of around 62 million grid cells.
All boundaries are considered as no-flow/no-slip boundaries, except at the left and right end of the free flow channel, where fixed pressure values
$p_\mathrm{in} = \SI{1e-3}{\pascal}$ and $p_\mathrm{out} = \SI{0}{\pascal}$ are set. The problem was solved using OpenFoam which took around five
hours on 30 cores (Intel Xeon CPU E5-2683 v4 @ 2.10GHz). Note that the chosen solver \texttt{icoFoam} (PISO algorithm) is implemented in a transient manner, thus the
simulation was run until a steady state was reached. Solving one time step took around four minutes.

\begin{figure}[H]
\centering
\begin{subfigure}[b]{0.04\linewidth}
\captionsetup{skip=40pt, justification=centering}
\caption{\label{fig:openfoam_v}}
\end{subfigure}
\begin{subfigure}[b]{0.95\linewidth}
\hspace*{18mm}\includegraphics[width=0.8\textwidth]{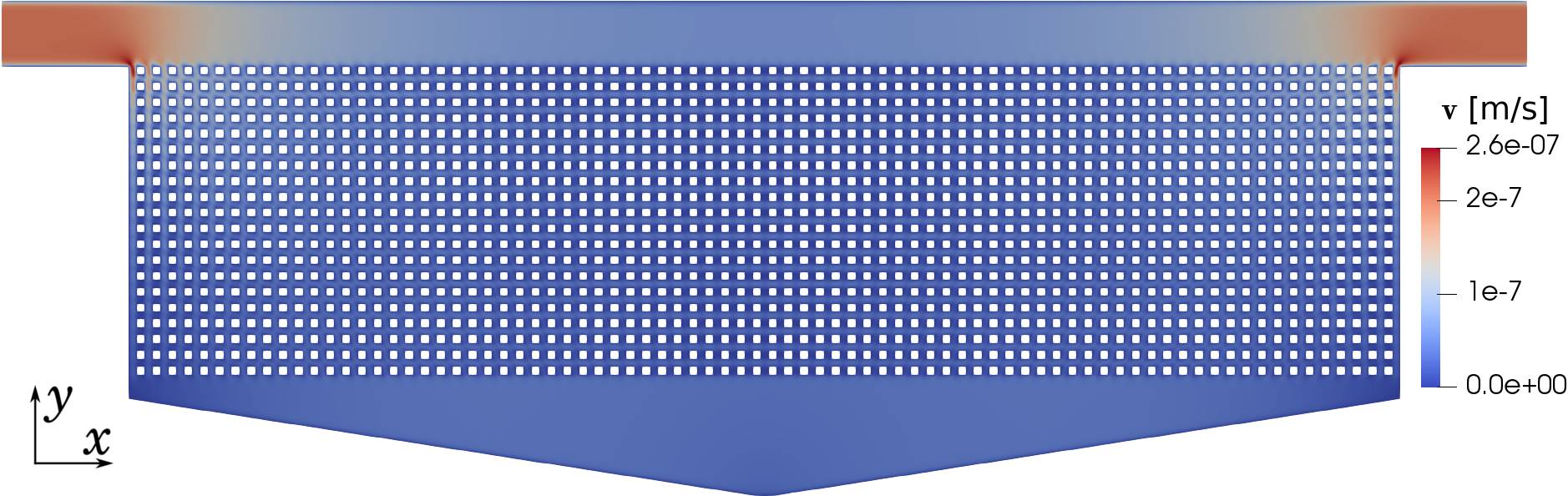}
\end{subfigure} \\[1ex]
\begin{subfigure}[b]{0.04\linewidth}
\captionsetup{skip=40pt, justification=centering}
\caption{\label{fig:openfoam_p}}
\end{subfigure}
\begin{subfigure}[b]{0.95\linewidth}
\hspace*{18mm}\includegraphics[width=0.8\textwidth]{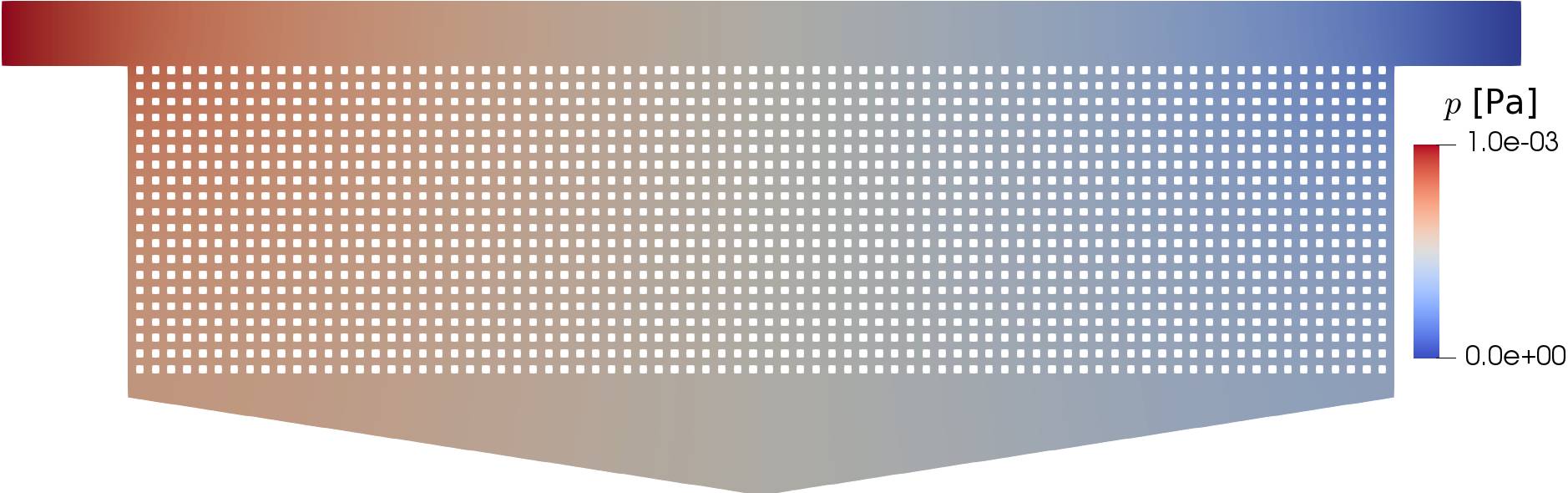}
\end{subfigure} \\[.1ex]
\begin{subfigure}[b]{0.04\linewidth}
\captionsetup{skip=70pt, justification=centering}
\caption{\label{fig:p_line}}
\end{subfigure}
\begin{subfigure}[b]{0.95\linewidth}
\includegraphics[width=1.0\textwidth]{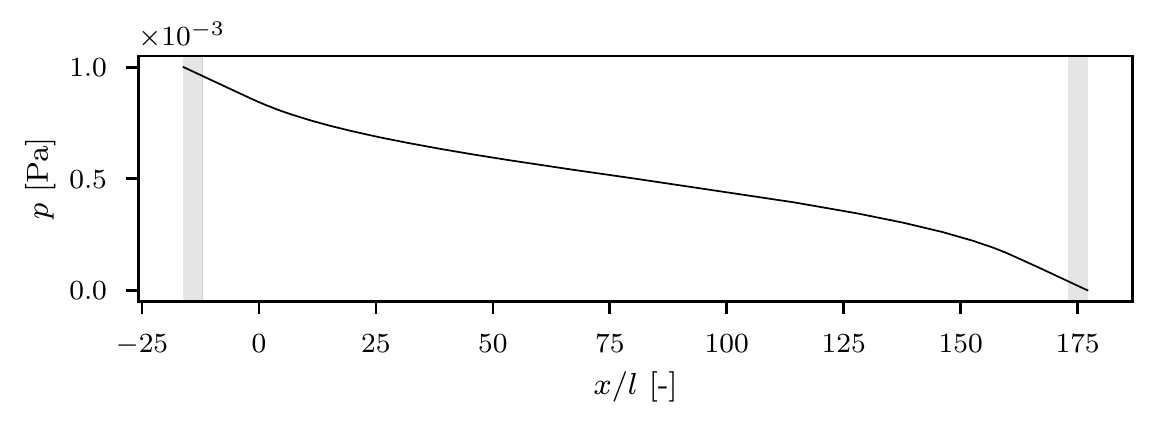}
\end{subfigure}
\caption{Center-plane velocity (a) and and pressure (b) field ($z= \SI{100e-6}{\meter}$) obtained by the 3D model (OpenFoam). (c) shows the pressure along the free-flow channel's central axis, the gray boxes indicate an almost linear pressure gradient. %
         Note that \texttt{icoFoam} returns $p*=p/\varrho$, depicted here is $p$.}
\label{fig:openfoam_fields}
\end{figure}

Fig. \ref{fig:openfoam_v} shows the resulting center-plane ($z= \SI{100e-6}{\meter}$) velocity field. As seen in the microfluidic experiments \citep{terzis2019a},
the flow enters the porous domain almost vertically at the left side of the porous medium, traverses it mainly parallel and re-enters the channel
at the right side of the porous domain. A substantial fraction of flow passes through the triangular reservoir at the bottom of the model as this features
less resistance than the narrow flow channels within the porous medium. The maximum resulting Reynolds number, both with respect to the free-flow channel width
and the width of the pore throats, is always below 0.001.

The corresponding pressure field of the entire domain is presented in Fig. \ref{fig:openfoam_p} while Fig. \ref{fig:p_line}
shows the pressure along the central axis of the free-flow channel. An approximately linear pressure gradient is found close to the inlet and the outlet (marked with gray boxes) as there is virtually no flow in $y$-direction.

\begin{figure}[H]
\centering
\includegraphics[width=0.3\textwidth]{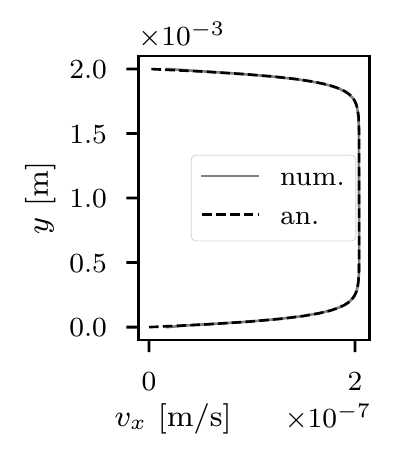}
\includegraphics[width=0.3\textwidth]{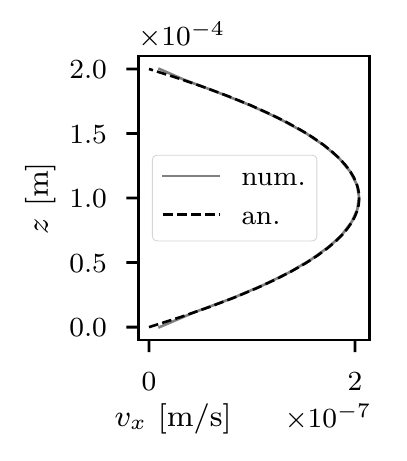}
\caption{Stream-wise velocity profiles at the left inlet of the free flow channel over the channel width ($y$, left) and the channel height ($z$, right) obtained by the 3D model %
        (OpenFoam). Both the numerical result and the analytical solution \citep{white2006a} for the given pressure gradient are shown.}
\label{fig:openfoam_analytical_v_y}
\end{figure}

We use this pressure gradient value to evaluate the analytical solution \citep{white2006a}
for the free flow channel's cross-sectional flow profiles at the inlet and assert that they are essentially identical to the corresponding numerical values as shown in
Fig \ref{fig:openfoam_analytical_v_y}. Using the same approach, we find that the integral volumetric flow trough the channel inlet determined numerically (see Eq. \eqref{eq:3d_flow_rate}) deviates
by less than \SI{0.3}{\percent} from the analytical solution \citep{white2006a}. The throats at $x/l \approx 80.5$
feature almost parallel flow along the $x$-axis where again a close to linear pressure gradient in $x$-direction can be
found. At $(x/l = 79.5, y/l = 37.5)$, which is close to the lower interface towards the triangular region, the evaluated pressure gradient within the throat yields an analytic flow rate from which the numerical results deviate
by less than \SI{1}{\percent}. This underlines that the mesh resolution is sufficiently fine.

\begin{figure}[H]
\centering
\includegraphics[width=0.4\textwidth]{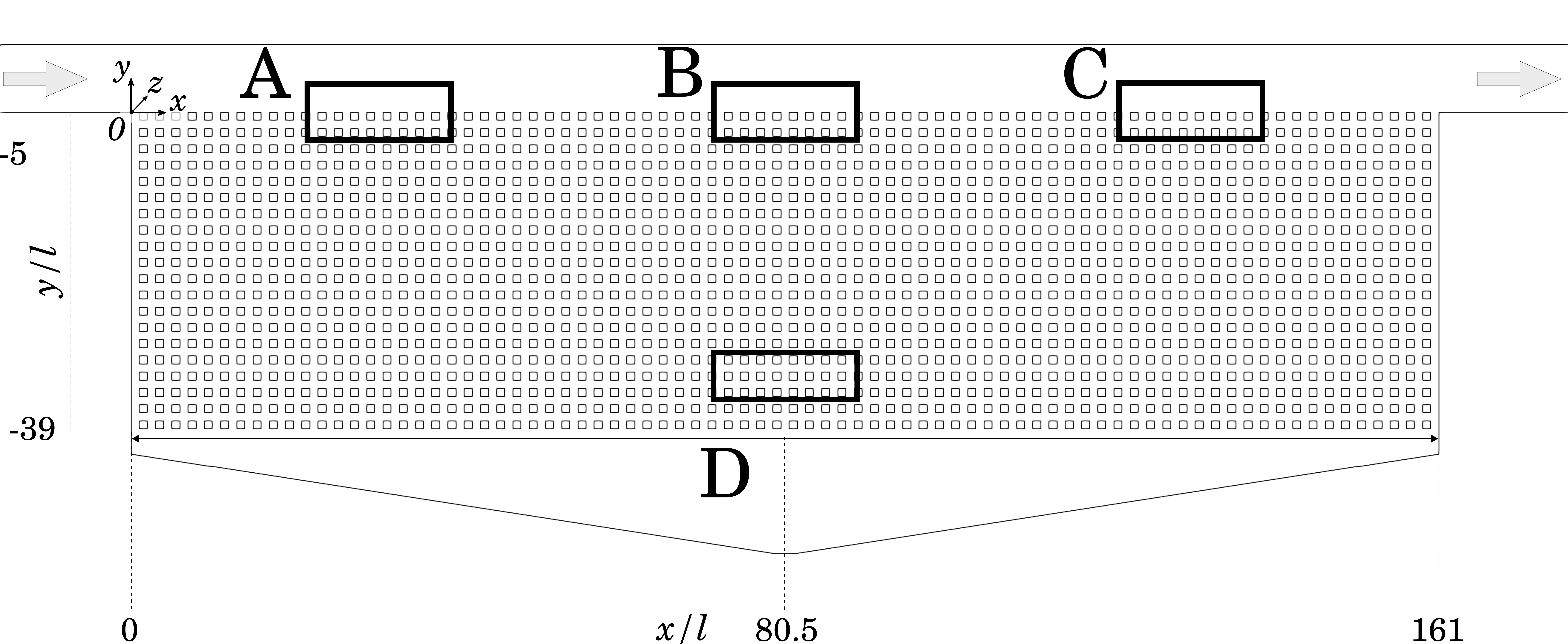} \\
\vspace*{0.2cm} \includegraphics[width=1.0\textwidth]{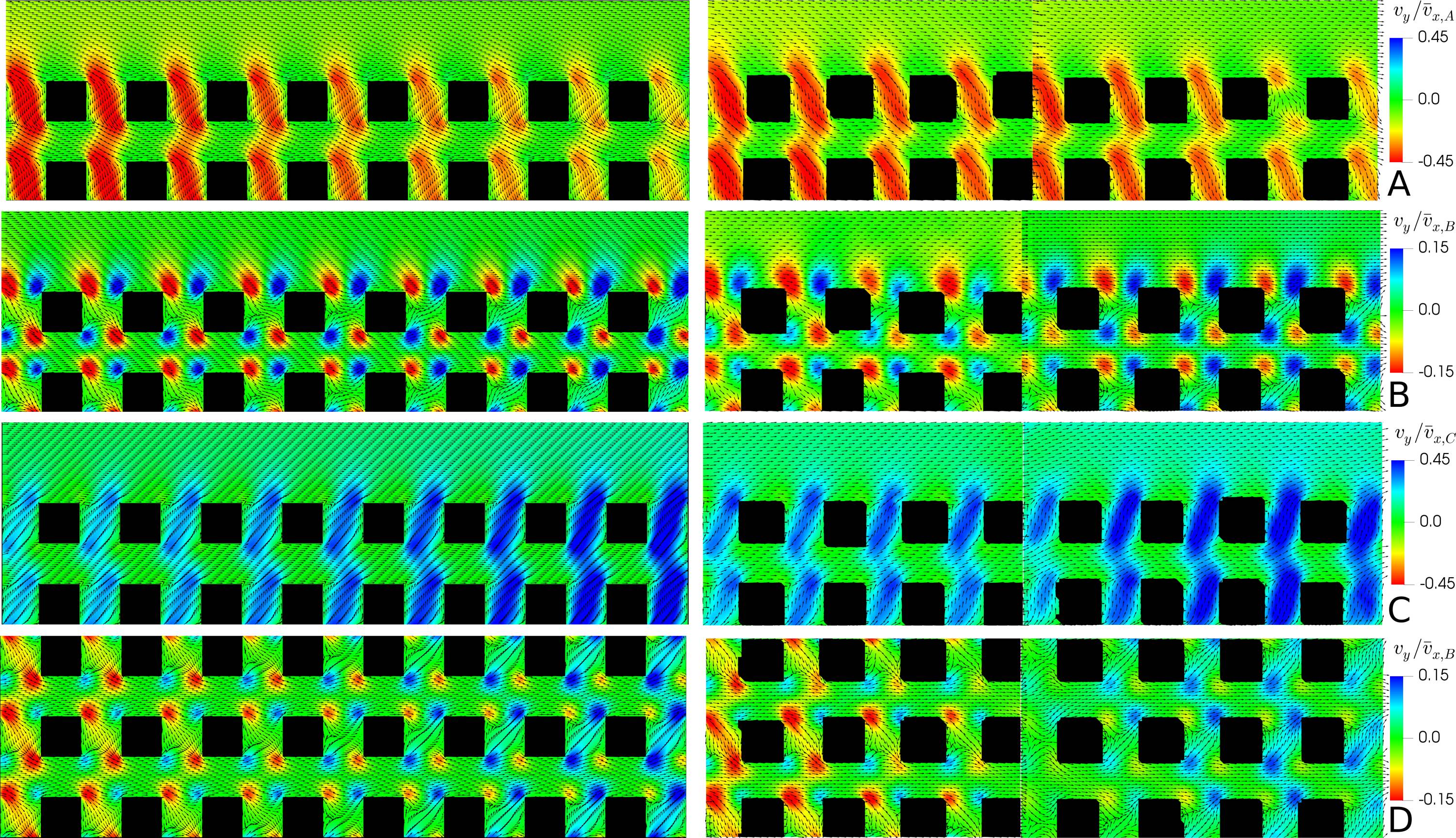}
\caption{Comparison of flow fields obtained by numerical simulation with OpenFoam (left column) and micro-PIV measurement results (right column, reproduced from the original data of \cite{terzis2019a}).}
\label{fig:fig_7_a}
\end{figure}

Fig. \ref{fig:fig_7_a} is a reproduction of Fig. 7a presented in \cite{terzis2019a}, using the same experimental data and color scheme. In the right column,
the results of the micro-PIV measurements are shown in the form of $v_y$ and the velocity vector fields for four different locations A, B, C and D as indicated in the schematic
drawing on the top of the figure. The left column displays the corresponding simulation results (OpenFoam) which show a very good agreement with the experimental data
in a qualitative sense, reproducing the same distinct flow patterns at the various locations: in region A, a pronounced inflow from the free flow channel into the
porous structure can be observed which diminishes in streamwise  direction. Region C basically shows a mirrored flow field as the fluid leaves the porous medium
and re-enters the channel in a symmetrical fashion compared to A. Here, the vertical flow intensity increases again in streamwise direction. In region B, at the
center of the porous domain, no net influx or outflux occurs. The fluid crosses the interface in a downwards motion at the right sides of the solid blocks (red spots)
and returns to the free flow channel at left sides of the blocks (blue spots). Region D lies inside the porous domain and features mainly parallel flow in $x$-direction.

\begin{figure}[H]
\centering
  \begin{subfigure}[b]{0.6\linewidth}
  \centering
    \includegraphics[width=1\textwidth]{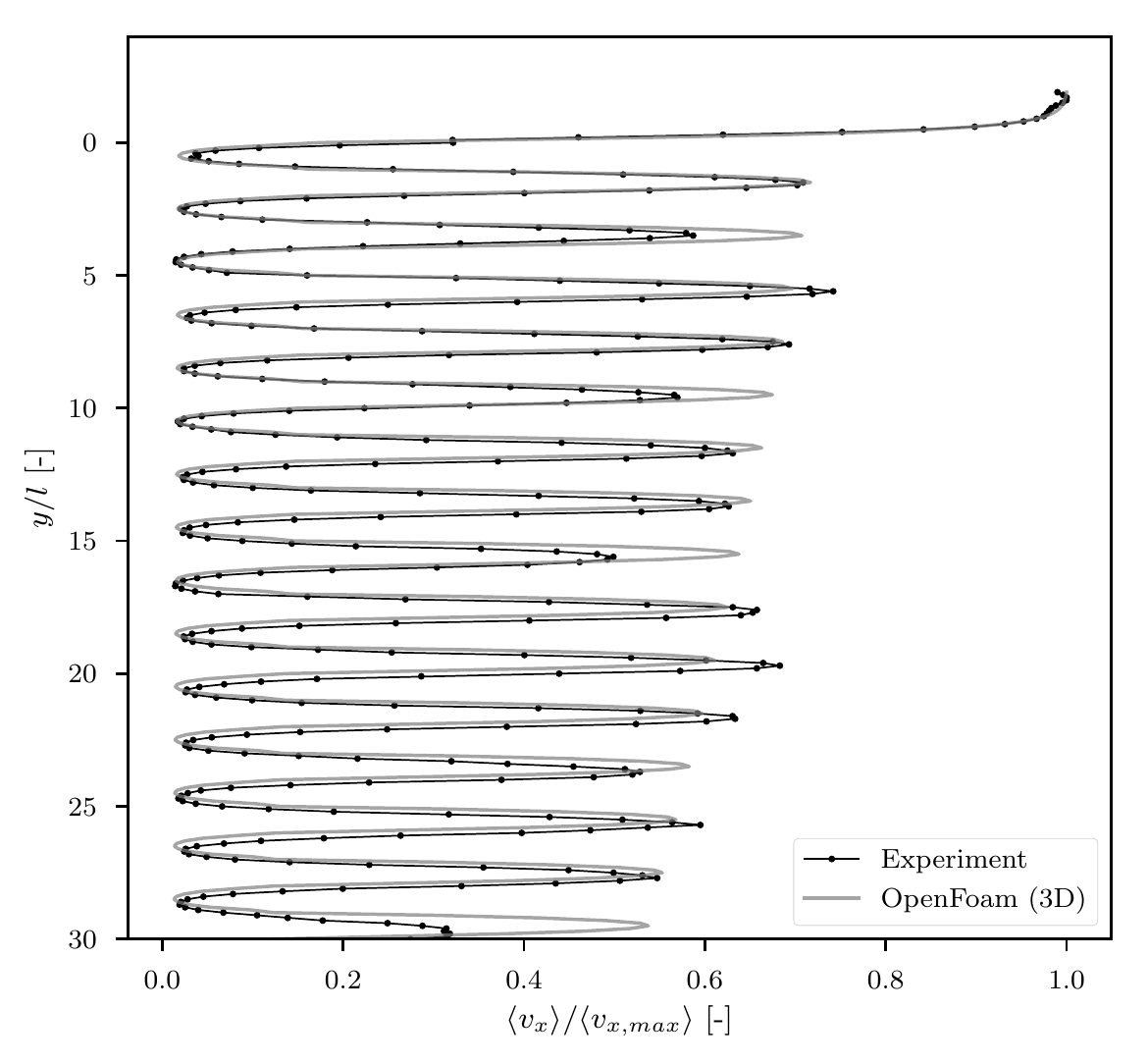}
    \caption{\label{fig:fig_9}}
  \end{subfigure} \hfill
  \begin{subfigure}[b]{0.3\linewidth}
    \centering
     \raisebox{15mm}{\includegraphics[width=1\textwidth]{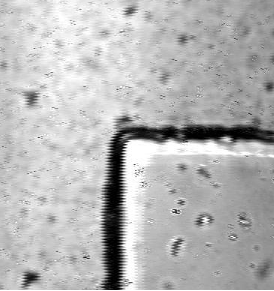}}
    \caption{\label{fig:foto_pillar}}
  \end{subfigure}
\caption{Left (a): Comparison of averaged velocity profiles ($75 \leq x/l \leq 85$) between simulation (OpenFoam) and experiment. %
         The original data of \cite{terzis2019a} were used. Right (b): camera image of a pillar within the porous domain.}
\label{fig:experiment}
\end{figure}

\begin{figure}[H]
\centering
\includegraphics[width=1.0\textwidth]{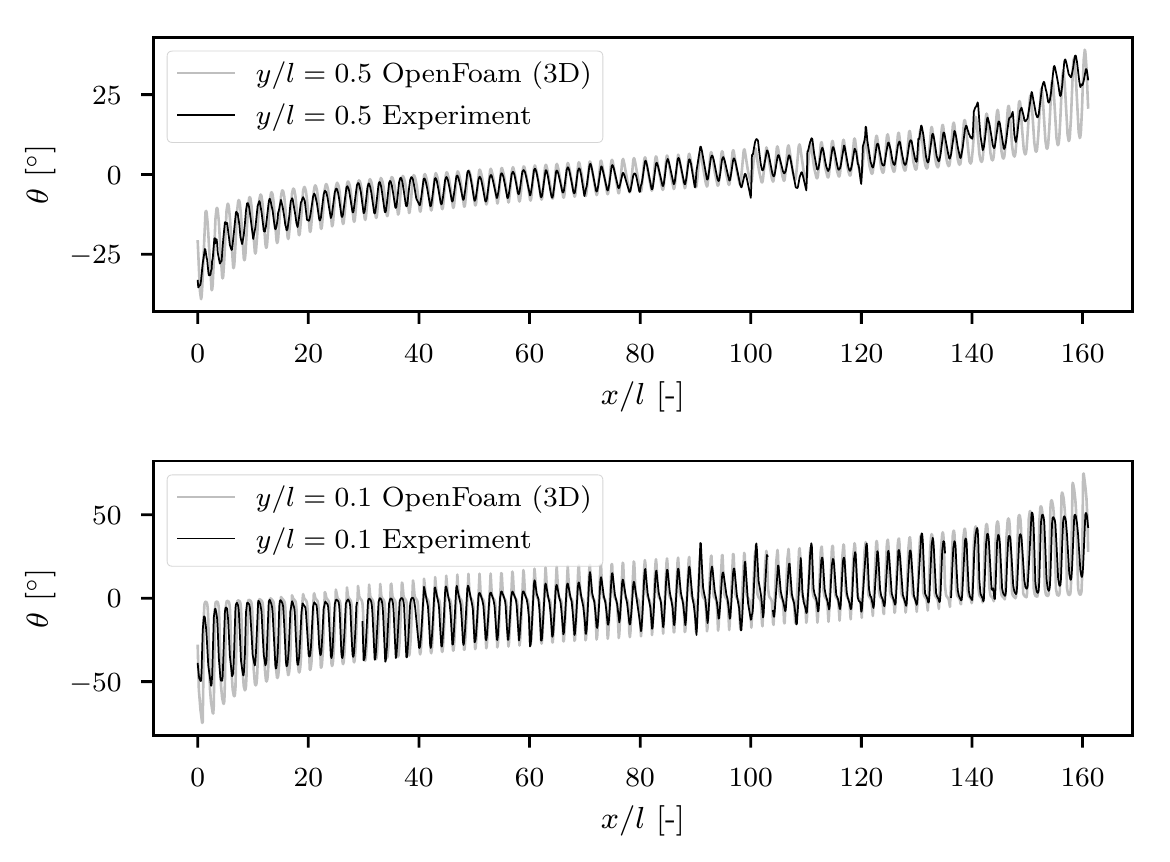}
\caption{Comparison of the flow angles $\theta$ close to the interface between free flow and porous medium at $y/l=0.5$ (top) and $y/l=0.1$ (bottom) for the simulation and the experiment.The original data of \cite{terzis2019a} were used.}
\label{fig:angles}
\end{figure}

In Fig. \ref{fig:fig_9}, a more quantitative comparison is performed. Here, $v_x$ is averaged in $x$-direction
between \mbox{$75 \leq x/l \leq 85$} at different locations of $y$. Both the experimental and numerical data are given
and the graphs are normalized by the respective maximum values in the free flow region. A very good fit can be found, both qualitatively and quantitatively.
The local deviations can be explained by measurements uncertainties \citep{terzis2019a} or small-scale structural differences between the actual micromodel geometry
and the computational domain, such as surface roughness \citep{silva2008a} which is not captured by the numerical model. While Fig. \ref{fig:foto_pillar} shows that the pillars of the PDMS
model are indeed not entirely smooth and that the corners are slightly rounded, the numerical model only considers perfectly smooth squares with sharp corners. This could also explain the local
deviations of flow angles $\theta$ close to the interface between the free-flow channel and porous medium, as presented in Fig. \ref{fig:angles}. A detailed analysis of the impact of the
pillars' rounded edges is beyond the scope of this paper and will be subject of future studies.

%\begin{figure}[H]
%\centering
%\includegraphics[width=0.7\textwidth]{./fig_9_center.pdf}
%\caption{Comparison of averaged velocity profiles ($75 \leq x/l \leq 85$) between simulation (OpenFoam) and experiment. The original data of \cite{terzis2019a} were used.}
%\label{fig:fig_9}
%\end{figure}

Fig. \ref{fig:angles} shows a symmetric characteristic of the flow angles due to the
inflow into the porous medium and the outflow back into the free flow channel. On the left,
the velocity vectors feature a negative inclination as the flow enters the porous domain while the same angles with opposed sign can be found on the right side,
where the flow returns to the free flow channel. The local oscillations are caused by the same up- and downwards movement of the flow between
the pillars as explained for region B in Fig. \ref{fig:fig_7_a}. The angles are greater for $y/l = 0.1$, which is closer to the interface, as the free flow in channel
senses a stronger influence of the porous medium compared to $y/l = 0.5$.

In summary, we find that the numerical model is able to reproduce the experimental data adequately. In the following section, the 3D simulation results will therefore serve
as a reference solution against which the reduced model will be compared.

\subsection{Quasi-3D simulation results}
\begin{figure}[H]
\centering
\includegraphics[width=0.8\textwidth]{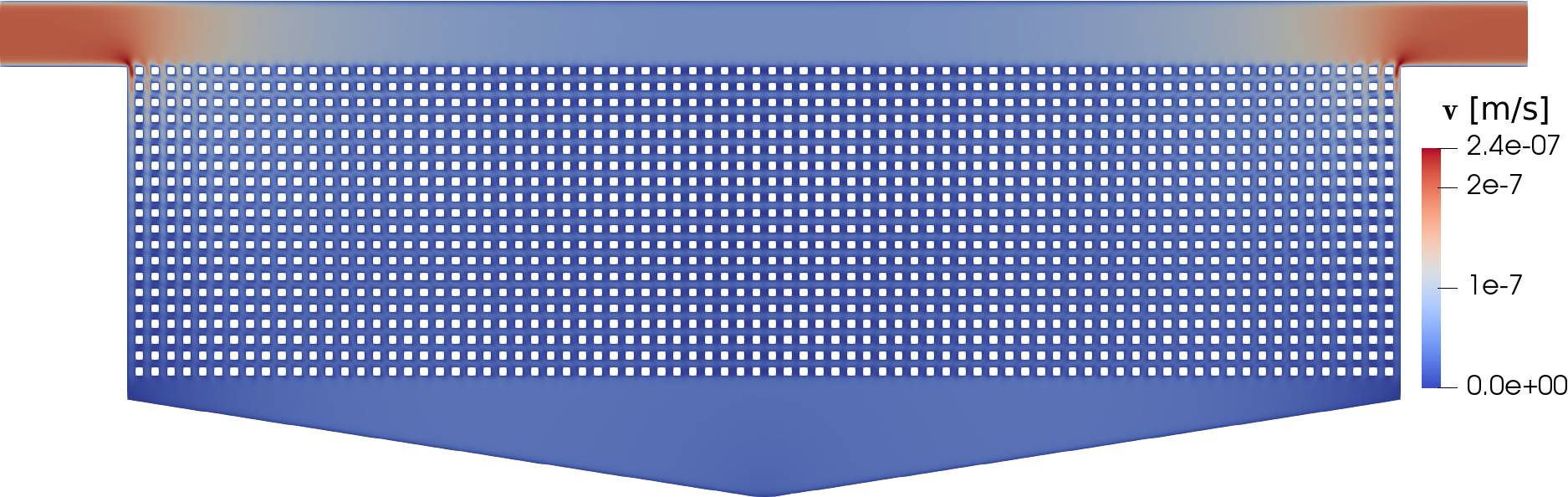}
\caption{Two-dimensional velocity field ($\mathbf{v}_\mathrm{2D}$) obtained by the quasi-3D model corresponding to the center plane ($z= \SI{100e-6}{\meter}$) of the 3D model.}
\label{fig:dumux_v}
\end{figure}

In order to assess the validity of the quasi-3D approach for the given micro-model geometry, the three-dimensional mesh used previously is flattened by
neglecting the $z$-coordinate which reduces the number of grid cells by a factor of 20. The same boundary conditions as before are applied, i.e, a pressure gradient
from left to right. Solving the quasi-3D problem with \dumux on a single core of the same machine as for the 3D model takes less than 11 minutes compared to five hours for the full 3D simulation. As mentioned earlier,
the current implementation of the quasi-3D model is based on \texttt{UMFPack} as direct linear solver which does not support parallelization. The extension to
preconditioned Krylov-type solvers is part of ongoing work. Fig. \ref{fig:dumux_v} shows the resulting velocity field which corresponds to the central $z$-plane of the three-dimensional model.

\begin{figure}[H]
\centering
\includegraphics[width=0.8\textwidth]{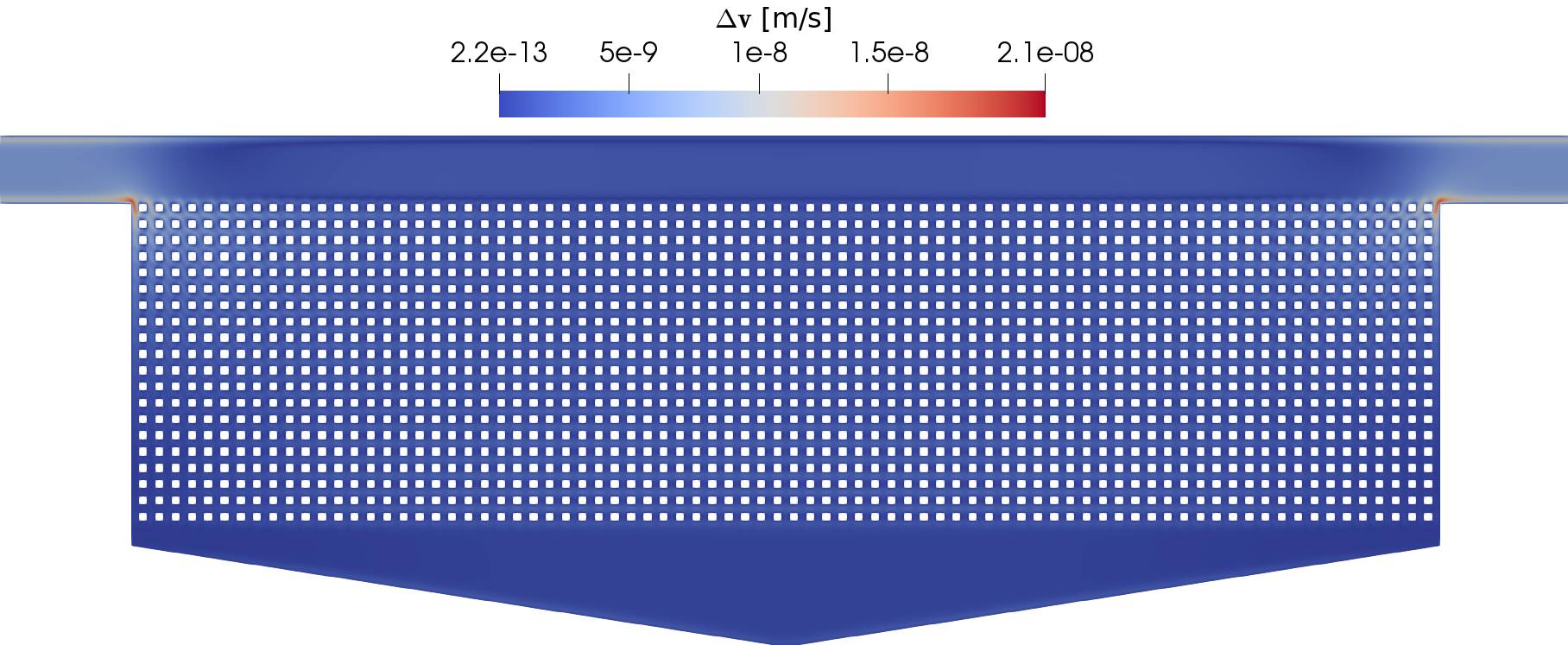}
\caption{Difference between the 3D and quasi-3D velocity fields.}
\label{fig:comparison_dumux_openfoam}
\end{figure}

The quasi-3D model captures the main features of the flow accurately when compared to Fig. \ref{fig:openfoam_v}. For a quantitative comparison,
Fig. \ref{fig:comparison_dumux_openfoam} shows the difference

\begin{equation}
\Delta \mathbf{v}_\mathrm{2D-3D} = \begin{pmatrix} v_{\mathrm{3D},x} - v_{\mathrm{2D},x} \\
                                                   v_{\mathrm{3D},y} - v_{\mathrm{2D},y}
                                  \end{pmatrix}
\end{equation}

between the two velocity fields.

Local deviations of up to $\SI{8}{\percent}$ can be observed, especially at the leftmost and rightmost vertical throat intersecting with the interface between free flow and porous medium. This is probably due the velocity gradients which are highest at these positions and the sudden change of flow direction. In addition, the aspect
ratio between the model height $h$ and the flow cross-section changes from a value of 0.1 ($\frac{\SI{200}{\micro \meter}}{\SI{2000}{\micro \meter}}$) in the channel to a less favorable value of 0.83
($\frac{\SI{200}{\micro \meter}}{\SI{240}{\micro \meter}}$) in the pore throats, which impairs the validity of Eq. \eqref{eq:flekkoy}. Globally, the deviations are quite small which corresponds to
the findings of \cite{venturoli2006a} and \cite{laleian2015a}. We define a relative error as the euclidean norm of the velocity differences normalized by the euclidean norm of the reference
velocities,

\begin{equation}
  \label{eq:norm}
  \mathrm{relErr}(\Delta \mathbf{v}, \mathbf{v}_\mathrm{ref} ) =  \frac{\lVert \Delta \mathbf{v} \rVert_2}{\lVert \mathbf{v}_\mathrm{ref} \rVert_2} =
                                    \frac{\left ( \sum_i (\Delta v_{x}^2 + \Delta v_{y}^2)_i \right )^{1/2}}
                                    {\left (\sum_i (v_{\mathrm{ref},x}^2 + v_{\mathrm{ref},y}^2)_i \right )^{1/2}} ~,
\end{equation}

which takes a value of \SI{3.28e-2}{} for the quasi-3D solution compared to the 3D center-plane solution ($\Delta \mathbf{v} = \Delta \mathbf{v}_\mathrm{2D-3D}$ and
$\mathbf{v}_\mathrm{ref} = \velthreed$). Analogously, the relative error for $p$ is \SI{3.7e-3}{}.

% \begin{equation}
%   \mathrm{relErr}_\mathrm{2D-3D} =  \frac{\lVert \Delta \mathbf{v}_\mathrm{2D-3D} \rVert_2}{\lVert \mathbf{v}_\mathrm{3D}^* \rVert_2} =
%                                     \frac{\left ( \sum_i (\Delta v_{\mathrm{2D-3D},x}^2 + \Delta v_{\mathrm{2D-3D},y}^2) \right )^{1/2}}
%                                     {\left (\sum_i v_{\mathrm{3D},x}^2 + v_{\mathrm{3D},y}^2 \right )^{1/2}} ~,
% \end{equation}

\begin{figure}[H]
\centering
\includegraphics[width=1.0\textwidth]{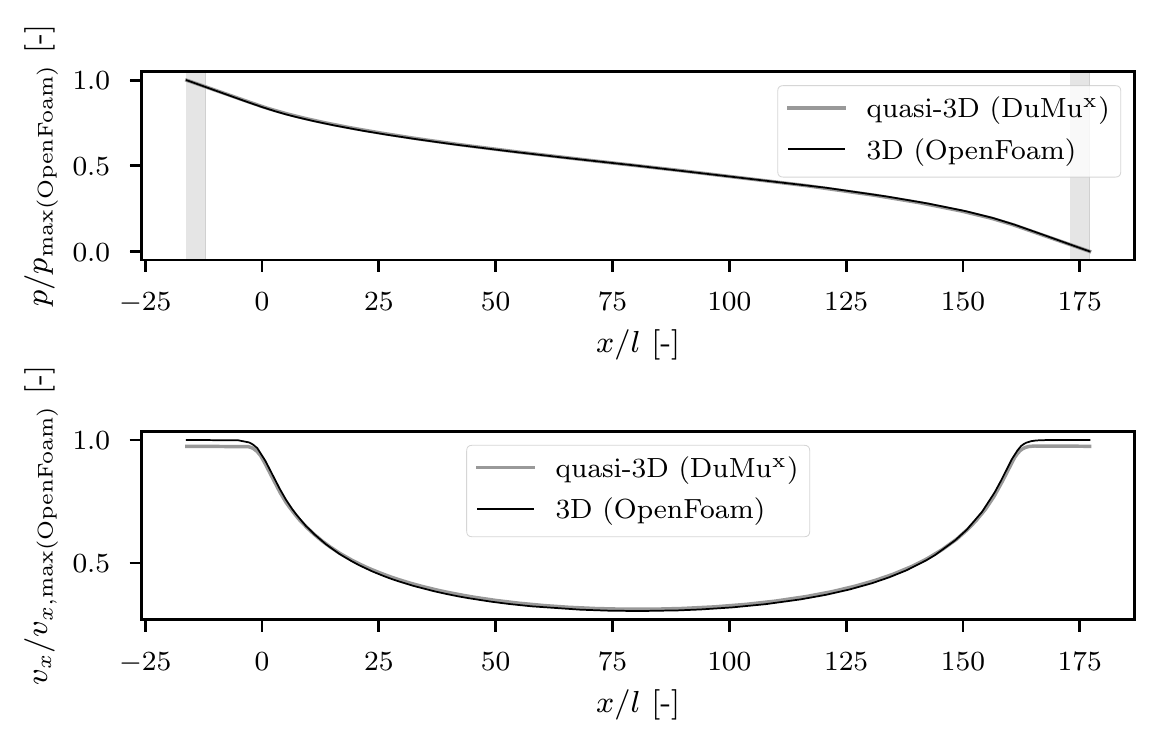}
\caption{Profiles of $p$ (top) and $v_x$ (bottom) along the central axis of the free-flow channel \mbox{($y=\SI{1e-3}{m}, z=\SI{100e-6}{m}$)} for both the 3D and quasi-3D model. The gray boxes indicate an almost linear pressure gradient. All results
         are normalized by the maximum values of the 3D model.}
\label{fig:comparison_dumux_openfoam_x}
\end{figure}

Fig. \ref{fig:comparison_dumux_openfoam_x} depicts the profiles of $p$ and $v_x$ along the free-flow channel's central axis in the $x$-direction for both the 3D and
quasi-3D simulation. All values are normalized by the maximum values of the 3D simulation. The pressure curves are virtually identical and, as stated earlier, a linear pressure
gradient can be found at the left part of the inlet channel as well as at the right part of the outlet channel. At these regions, some differences with regard to $v_x$ occur
($\SI{2.6}{\percent}$) while there is a very close match between the solutions otherwise.

\begin{figure}[H]
\centering
  \begin{subfigure}[b]{0.49\linewidth}
  \centering
    \includegraphics[width=1\textwidth]{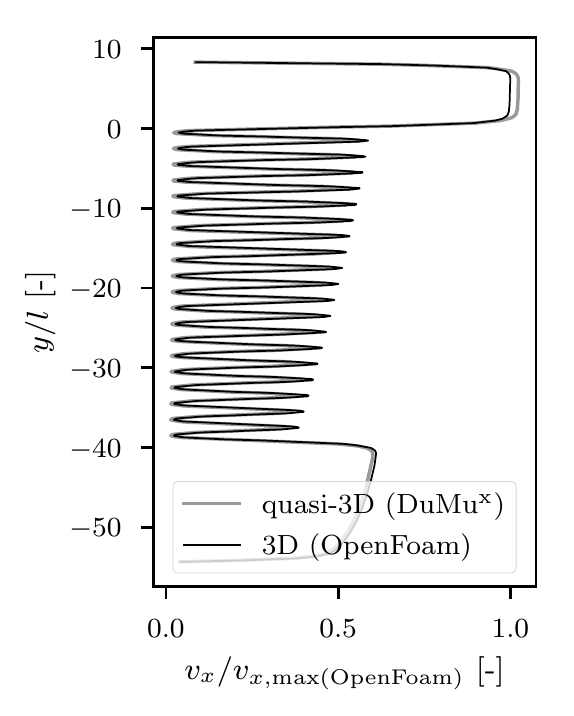}
    \caption{\label{fig:comparison_dumux_analytical_y_a}}
  \end{subfigure}
  \begin{subfigure}[b]{0.49\linewidth}
    \centering
    \raisebox{20mm}{\includegraphics[width=1\textwidth]{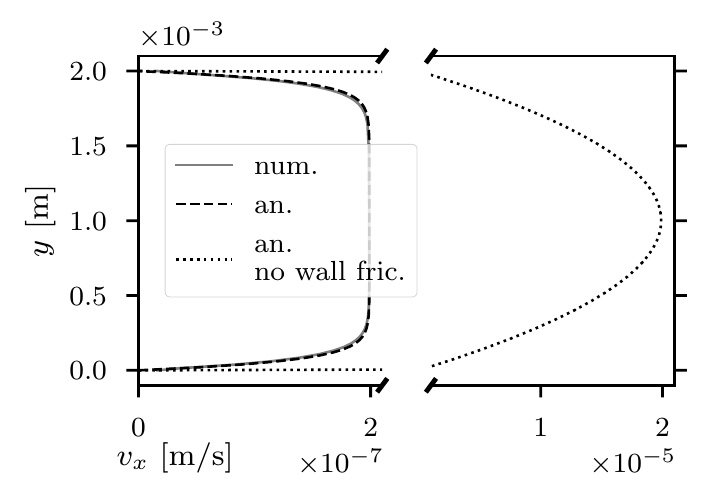}}
    \caption{\label{fig:comparison_dumux_analytical_y_b}}
  \end{subfigure}
\caption{Left (a): Velocity profiles $v_x$ over $y$ at $y/l = 80.5$ for the 3D and quasi-3D model. Right (b): Velocity profile at the inlet of the free-flow channel and corresponding %
        analytical solution \citep{white2006a}. For comparison, also a parabolic flow profile, neglecting the friction of the top and bottom wall is shown.}
\label{fig:comparison_dumux_analytical_y}
\end{figure}

The profiles of $v_x$ along the $y$-axis at the center of the micromodel \mbox{($x/l=80.5$)} are presented in Fig. \ref{fig:comparison_dumux_analytical_y_a}. Again, all values are
normalized by the maximum velocity obtained by the 3D model. The three flow domains of the micromodel can be distinguished clearly: first, the free-flow channel at the top ($y/l > 0$),
where there is a rather uniform flow profile, slightly skewed due to the interaction with the porous domain. The latter is seen at $-40 \leq y/l < 0$. Here the velocity peaks
between the solid blocks are distinctly visible while $v_x = 0$ at the locations of the blocks. There is also a gradual decrease of the peaks from top to bottom. The velocity
increases again in the triangular reservoir ($y/l < -40 $) where it reaches around $\SI{60}{\percent}$ of the free-flow channel velocity. The solution of the quasi-3D model
closely follows the reference solution. It slightly over-predicts $v_x$ with a maximum deviation of $\SI{7.6}{\percent}$ in the lower part of the porous medium. As previously mentioned,
the error within the pore throats is likely to be higher due to the less favorable aspect ratio of the geometry with respect to the accuracy of Eq. \eqref{eq:flekkoy}.

Focusing on the inlet part of the free-flow channel, we evaluate the local linear pressure gradient and use this value to compute the corresponding analytical solution for $v_x$ over
the channel width (see Fig. \ref{fig:comparison_dumux_analytical_y_b}, in analogy to Fig. \ref{fig:openfoam_analytical_v_y}). As before, the analytical and numerical solution are
virtually identical. For comparison, Fig. \ref{fig:comparison_dumux_analytical_y_b} also shows the analytical solution when not including the wall friction, which is just an
ordinary parabolic profile where the maximum velocity is two orders of magnitude higher. This highlights the importance of including Eq. \eqref{eq:flekkoy} for an accurate description of the
flow field. The integral volumetric flow based on Eq. \eqref{eq:2d_flow_rate} deviates by less than $\SI{0.9}{\percent}$ from the respective analytical value \citep{white2006a}. \\

\begin{figure}[H]
\centering
\includegraphics[width=1.0\textwidth]{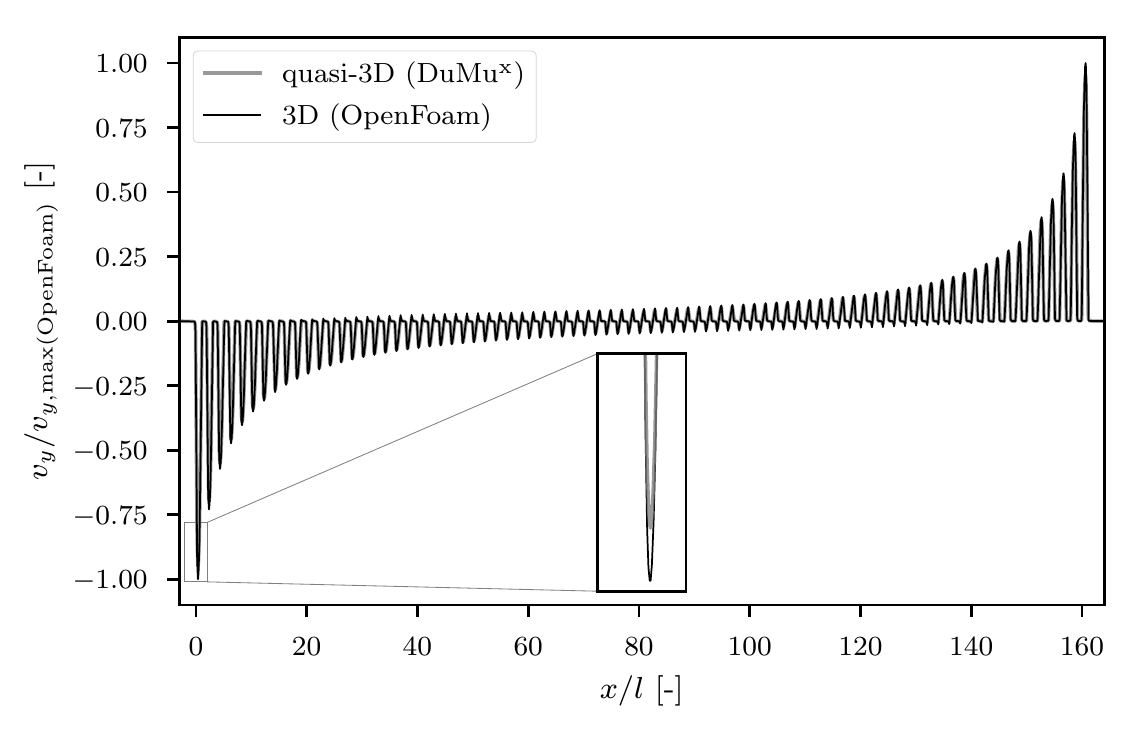}
\caption{Normalized $v_y$ at the interface in stream-wise direction ($y/l=0$) for the 3D and quasi-3D model.}
\label{fig:comparison_dumux_openfoam_interface}
\end{figure}

Fig. \ref{fig:comparison_dumux_openfoam_interface} shows the evolution of the vertical velocities $v_y$ along the interface for both the 3D and quasi-3D model.
As expected, there are peaks at the centers of the pore throats while $v_y$ goes to zero at the solid blocks. The highest inflow into the porous domain, i.e., the most negative $v_y$, occurs
at the very left throat. The amplitude of the velocity fluctuation then decreases until it reaches minimum at the center of the porous part ($x/l = 80.5$) after which it increases again in a
symmetric manner. The same behavior could be observed in the experiment \citep{terzis2019a} and in the simulations presented in \citep{weishaupt2019a}. Due to model symmetry,
$v_y = 0$ at $x/l = 80.5$. There is again a very good fit between the quasi-3D and the reference solution with the highest deviation at the leftmost and rightmost throat (\SI{6.5}{\percent}).

% \begin{figure}[H]
% \centering
% \includegraphics[width=0.6\textwidth]{./analytical_v_y_dumux.pdf}
% \caption{text}
% \label{fig:comparison_dumux_openfoam_analytical}
% \end{figure}

\subsection{Numerical determination of the conductance factor $k_{ij}$ and the slip coefficient $\beta_\mathrm{throat}$}
\label{sec:upscaling}
In the final step of model reduction, a pore-network model is used to account for the porous part of flow domain.
Here we describe and discuss the numerical evaluation of the input parameters required by the coupled hybrid-dimensional model.
First, effective conductance factors $k_{ij}$ need to be assigned to the pore throats of the pore network model. As shown in \cite{weishaupt2019a},
the given porous geometry, i.e., pore bodies and pore throats have the same dimensions, necessitates the inclusion of the pressure drops both within the pore bodies and
throats (e.g., \cite{dillard2000a, raoof2012a, mehmani2017a}):

\begin{equation}
\label{eq:sum_conductance}
k_{ij} = \left(k_{ij,t}^{-1} + k_{1/2, i}^{-1} + k_{1/2, j}^{-1} \right)^{-1} ~.
\end{equation}

The conductance factor of the throat itself is given by $k_{ij,t}$, those of the two adjacent pore body halves by $k_{1/2, i}$ and $k_{1/2, j}$, respectively.
Pore bodies intersecting with the interface are assumed to be volumeless and thus feature no resistance. The procedure of estimating the individual
conductance factors is explained and discussed extensively in the appendix of \cite{weishaupt2019a}: a pressure boundary value problem is solved numerically on a discretely resolved
(2D), reduced but equivalent structure of pillars of the same dimensions as in the micromodel in order to relate the pressure drop within the pore throats and bodies to the
resulting volume flows.

\begin{table}[H]
    \caption{Setups used for the numerical determination of $\beta_\mathrm{throat}$ for throat widths of $\SI{50}{\micro \meter}, \SI{100}{\micro \meter}, %
             \SI{200}{\micro \meter}$ and $\SI{400}{\micro \meter}$, as indicated by the white numbers. The throats are quadratic $w_\mathrm{throat} = l_\mathrm{throat}$, %
             as well as the free flow channel above them which features side lengths of $\SI{1e-3}{m}$. Depicted are the reference solution (left), the solution of the coupled model, %
             and a plot comparing the velocity profiles between the two models over the vertical white line. On the right, the values of $\beta_\mathrm{throat}$ are given. %
             The green boxes at the coupled model symbolize the virtual extent of the one-dimensional throats. Both the reference and coupled problems are solve with \dumux.}
    \label{tab:slip}
    \centering
    \begin{tabular}{c*{4}{c}}
        \hline
        \textbf{setup (reference / coupled model)}  & $v_x$ over $y$ &  $\beta_\mathrm{throat}$ [1/m]     \\
        \hline
        \rule{0pt}{6ex}  \begin{minipage}{.51\textwidth} \vspace*{-3.5mm}\includegraphics[width=\linewidth]{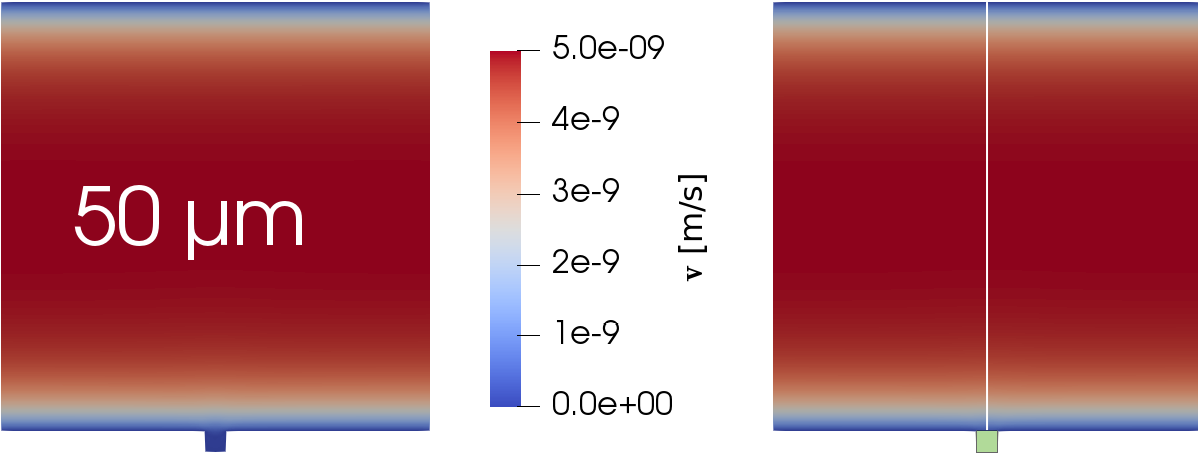} \end{minipage} &
                         \begin{minipage}{.25\textwidth} \vspace*{1mm}\includegraphics[width=1.05\linewidth]{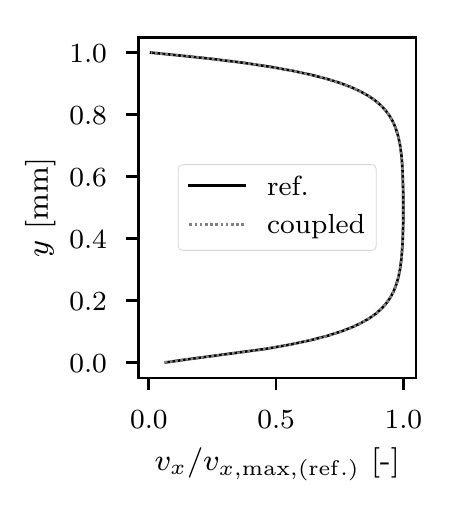} \end{minipage}  & 176416 \\
        \rule{0pt}{6ex}  \begin{minipage}{.51\textwidth} \vspace*{-5mm}\includegraphics[width=\linewidth]{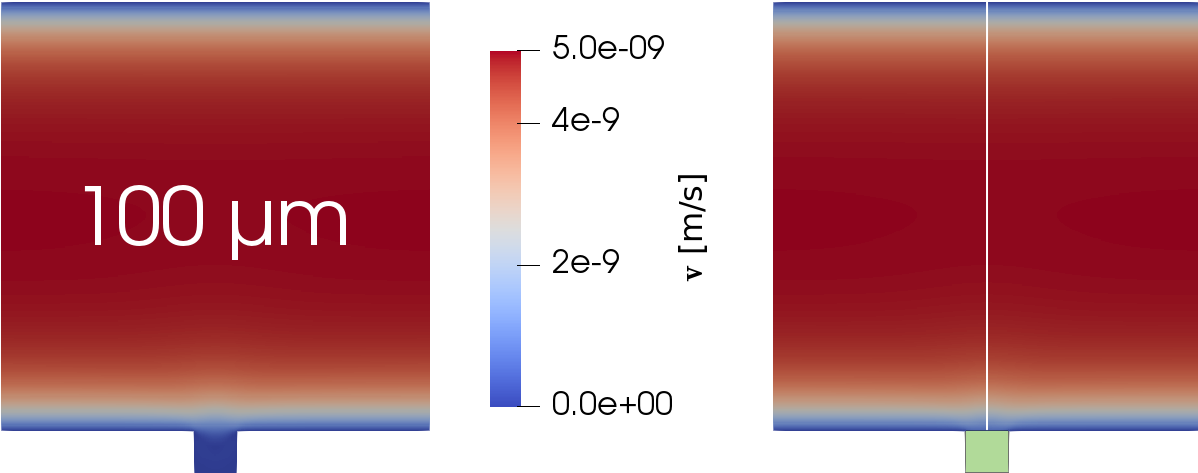} \end{minipage} &
                         \begin{minipage}{.25\textwidth} \includegraphics[width=1.05\linewidth]{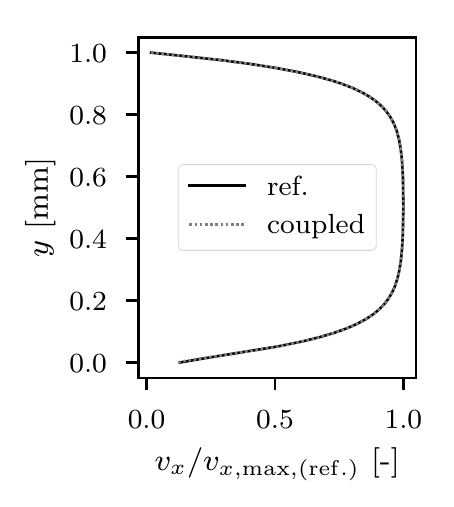} \end{minipage}  & 84550 \\
        \rule{0pt}{6ex}  \begin{minipage}{.51\textwidth} \vspace*{-2.5mm}\includegraphics[width=\linewidth]{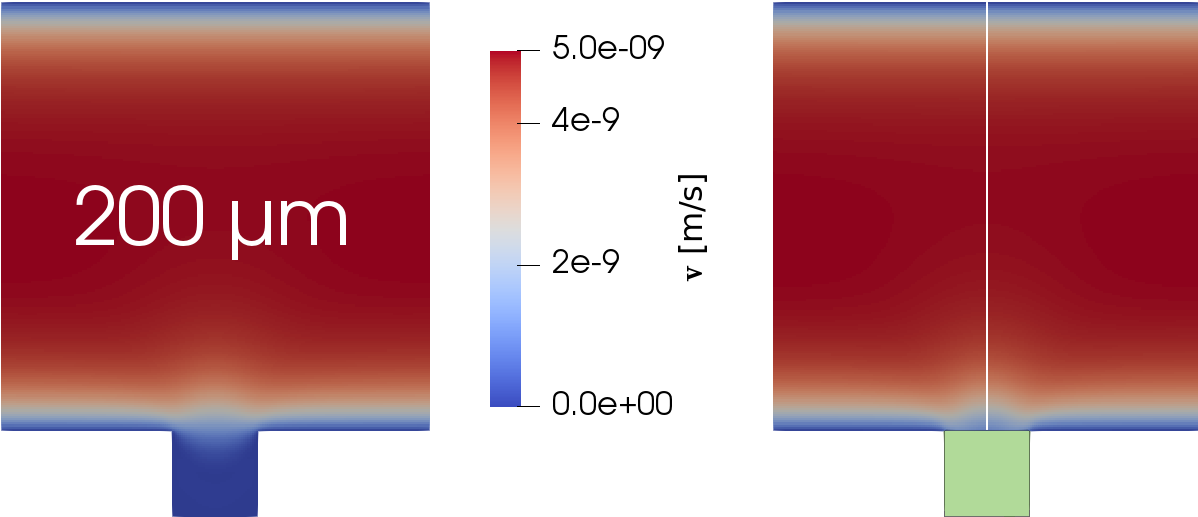} \end{minipage} &
                         \begin{minipage}{.25\textwidth} \includegraphics[width=1.05\linewidth]{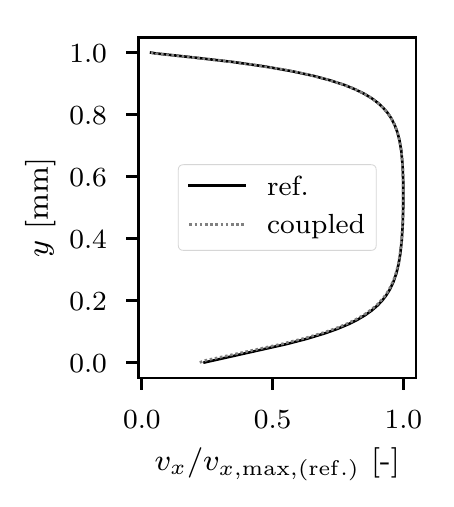} \end{minipage}  & 39924 \\
        \rule{0pt}{6ex}  \begin{minipage}{.51\textwidth} \vspace*{2.5mm}\includegraphics[width=\linewidth]{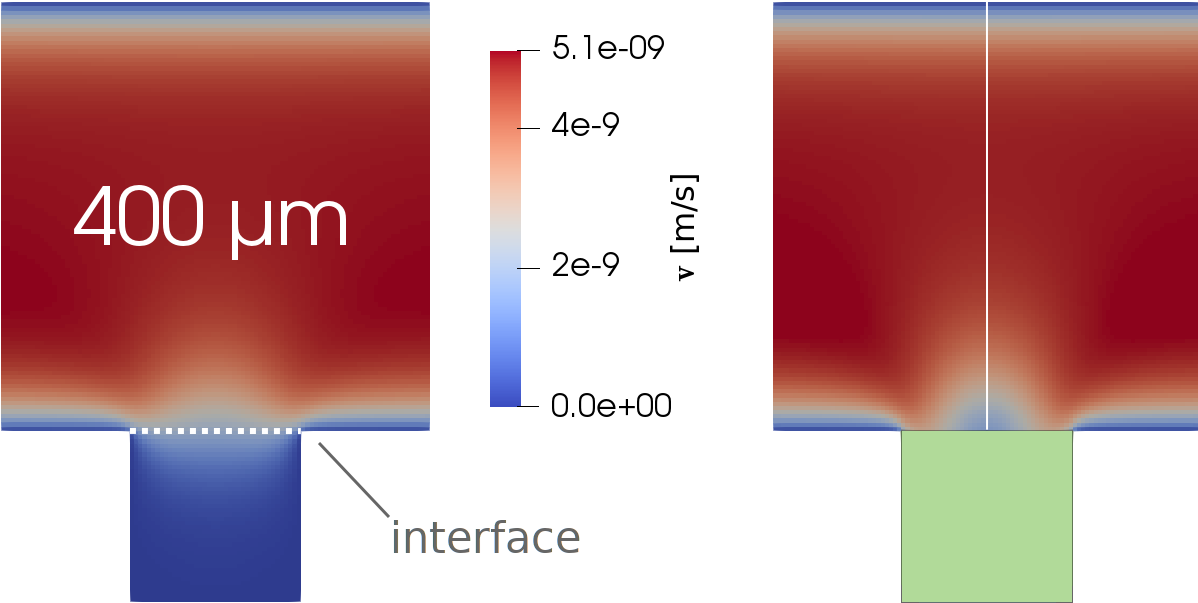} \end{minipage} &
                         \begin{minipage}{.25\textwidth} \includegraphics[width=1.05\linewidth]{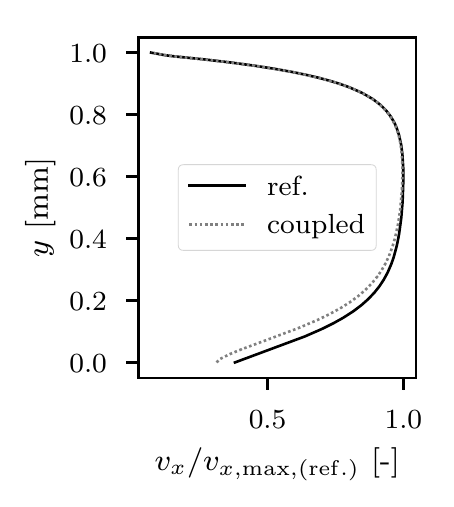} \end{minipage}  & 19087 \\
        \hline
    \end{tabular}
\end{table}
\newpage

From this, the resistance factors can be recalculated. Following this approach yields $k_{ij,t} \approx \SI{3.05e-10}{m^3/(s \pascal)}$ and
$k_{1/2, i} = k_{1/2, j} \approx \SI{8.47e-10}{m^3/(s \pascal)}$. \\

The second input parameter needed for the coupled model is the pore-scale slip coefficient $\beta_\mathrm{throat}$. We approximate this value by solving a simplified,
equivalent problem of free flow over a single pore throat intersecting with the lower boundary of the free-flow channel (see dotted line at the
bottom of Tab. \ref{tab:slip}). We considered four different throat widths to find a relation between $\beta_\mathrm{throat}$ and the
throat width $w_\mathrm{throat}$ which could be used for a wider range of different porous geometries.
For all cases, a quadratic free-flow channel with a side length of $\SI{1e-3}{m}$ and a virtual height $h = \SI{200}{\micro \meter}$ is considered.
The throats are also quadratic and feature the same $h$. A pressure drop from left to right $\Delta p_\mathrm{in} = \SI{1e-9}{\Pa}$ is assigned as boundary condition,
while the remaining boundaries have no flow/no slip conditions. Like in the calculation of the conductance described above, the geometry is meshed discretely
in 2D and the quasi-3D Stokes model implemented in \dumux is employed. The slip velocities and velocity gradients at the interface between pore throat and free flow (corresponding to $y/l = 0$) are then
extracted, averaged and used to approximate the slip coefficient:

\begin{equation}
  \label{eq:beta_throat}
  \beta_\mathrm{throat} \approx \frac{ \left \langle \frac{\partial v_x }{\partial y} + \frac{\partial v_y }{\partial x} \right \rangle }  {\langle v_x \rangle }
\end{equation}

Given by the geometry, we neglect the horizontal velocity component in the interior of the pore throat (which would be zero in the pore-network model).
The resulting factors are given in the last column of Tab. \ref{tab:slip} and then used for recalculating the given setups using the coupled model, i.e, replacing the small
cavity by a single one-dimensional pore throat of the same width and length. The velocity fields of the reference (quasi-3D) and the coupled model are shown in the first
and second column of Tab. \ref{tab:slip}, next to a comparison of the velocity profiles. The latter shows virtually identical results for
\mbox{$w_\mathrm{throat} = \{50, 100, 200\} ~ \si{\micro \meter}$} while the coupled models slightly underestimates the slip velocities at the bottom of the
channel for $w_\mathrm{throat} = \SI{400}{\micro \meter}$. The smaller the throat width, the better the fit.

\begin{figure}[H]
\centering
\includegraphics[width=1.0\textwidth]{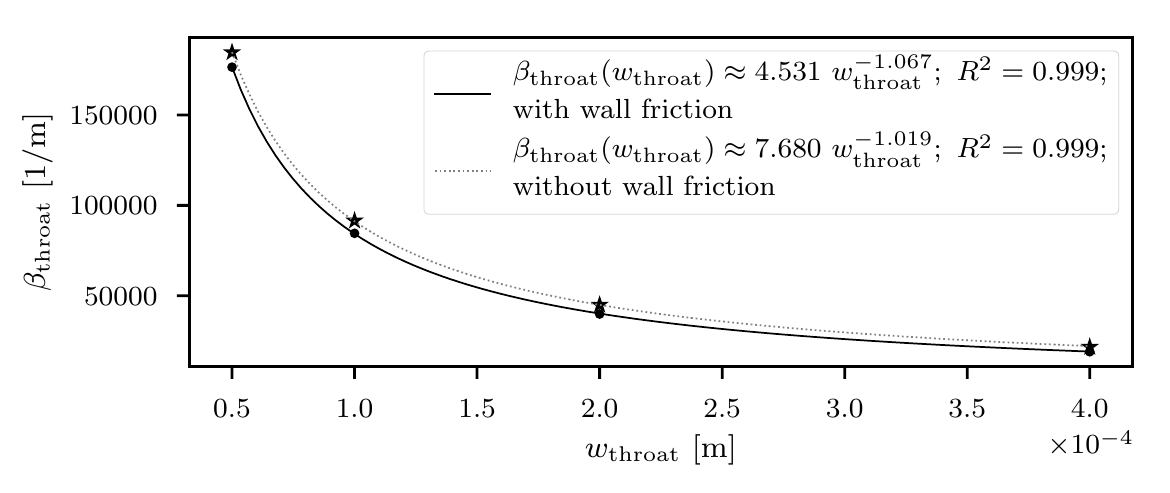}
\caption{Plot of $\beta_\mathrm{throat}$ over $w_\mathrm{throat}$ for a case considering the bottom and wall friction \mbox{(h = $\SI{200}{\micro \meter})$} and %
         a case neglecting this influence.}
\label{fig:betas}
\end{figure}

Fig. \ref{fig:betas} shows the different slip coefficients over $w_\mathrm{throat}$. For sake of completeness,
also values for purely two-dimensional setups neglecting the flow resistance of the bottom and top wall ($z$-coordinate) of the channel are shown. Using a power-law fit,
a functional relation between $\beta_\mathrm{throat}$ and $w_\mathrm{throat}$ is found empirically and given in Fig. \ref{fig:betas}. \\

\begin{figure}[H]
\centering
\includegraphics[width=0.7\textwidth]{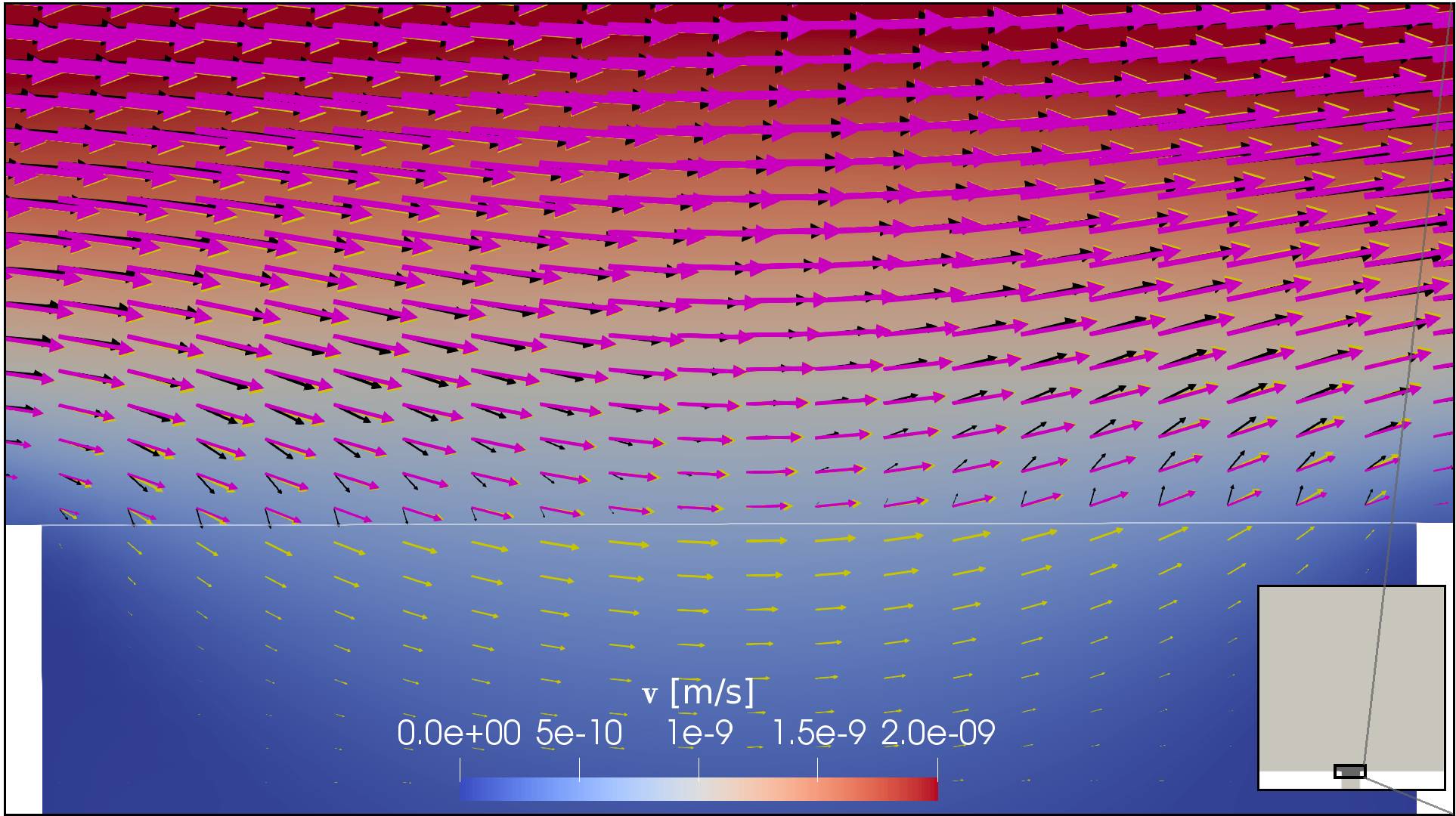}
\caption{Close-up of the interface region (see gray inlay) for a vertical throat with \mbox{$w_\mathrm{throat} = \SI{100}{\micro \meter}$}. %
         The yellow, purple and black velocity  vectors correspond to the reference (quasi-3D) model, the coupled model including the slip %
         term and the coupled model with a no-slip condition at the interface.}
\label{fig:orthogonal_slip}
\end{figure}

The validity of the novel approach for incorporating the slip velocity over pore throats at the interface in the coupled model is further assessed.
Fig. \ref{fig:orthogonal_slip} shows a close-up of the reference solution (quasi-3D model) for $w_\mathrm{throat} = \SI{100}{\micro \meter}$  at the interface region.
The yellow velocity vectors correspond to the solution of the reference model, the purple vectors to the solution of the coupled model including the throat slip coefficient.
For comparison, also black velocity vectors corresponding to the solution of the coupled model without considering a slip velocity at the interface ($v_x = 0$) are shown.
Note that the vectors of the coupled models are only given in the free-flow channel because the cavity (below the white line) is modeled as a one-dimensional throat where only an
averaged vertical velocity is defined. There is an excellent fit between the reference solution and the coupled model including the slip term, the corresponding velocity vectors
both feature very similar magnitudes and orientations. This is in contrast to the coupled model employing a no-slip condition at the interface. Here, the black velocity vectors
clearly deviate from the reference solution especially at the very left and very right part of the interface. The greater the distance to the interface, the smaller the deviations. The relative error for the velocities as defined in Eq. \eqref{eq:norm} reduces by a factor of 9 from \SI{5.26e-3}{} to \SI{5.83e-04}{} when including the slip above the throats. For the pressure, this error drops by a factor of 3.72 from \SI{1.85e-03}{} to \SI{4.98e-04}{}.

\begin{figure}[H]
\centering
\includegraphics[width=0.7\textwidth]{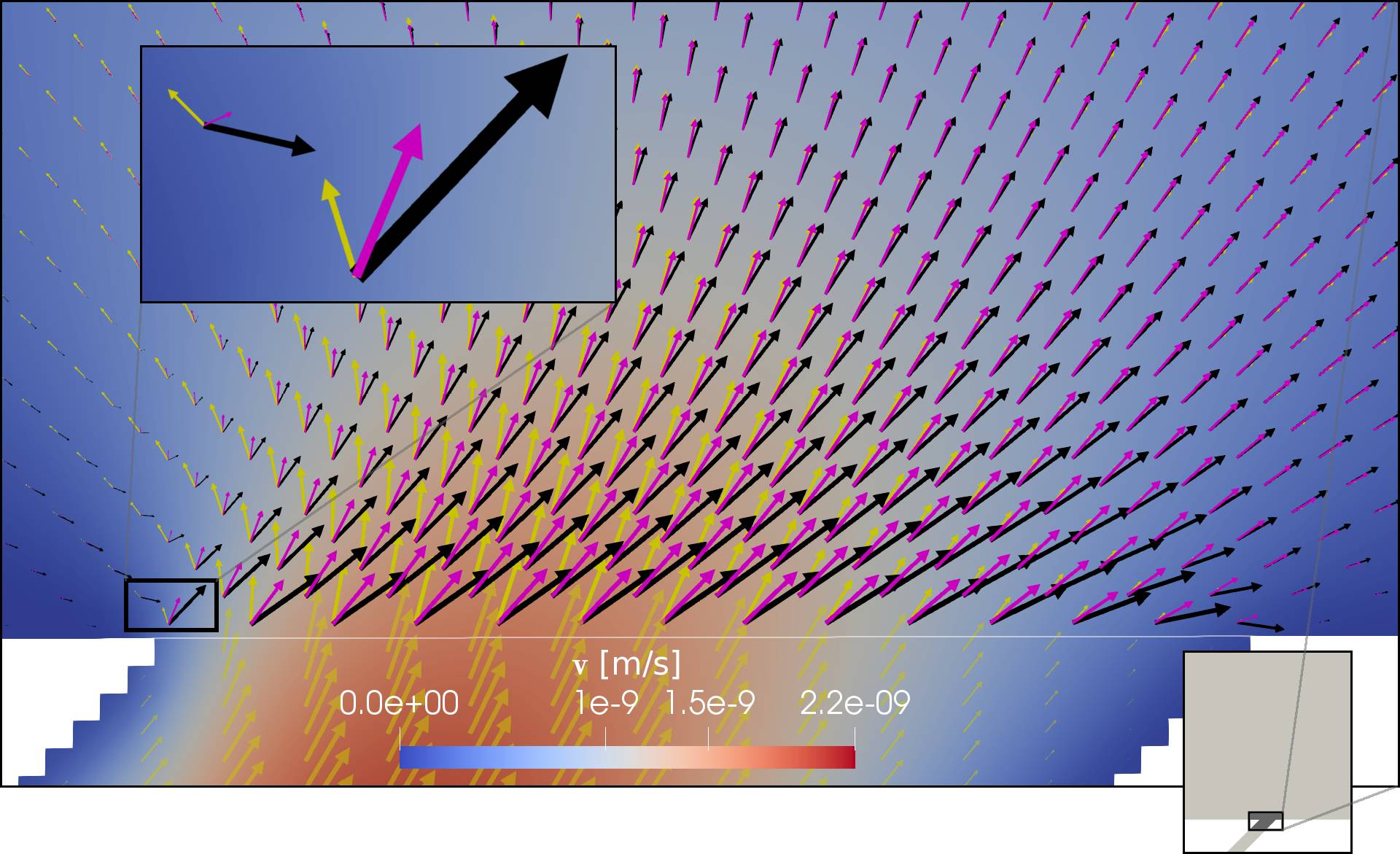}
\caption{Close-up of the interface region (see gray inlay) for an inclined throat with \mbox{$w_\mathrm{throat} = \SI{100}{\micro \meter}$}. %
         There is an inflow from the bottom of the throat and flow from left to right in the free-flow channel. %
         The yellow, purple and black velocity  vectors correspond to the reference (quasi-3D) model, the coupled model including the slip %
         term and the coupled model with a no-slip condition at the interface.}
\label{fig:skewed_slip_with_dp}
\end{figure}

Fig. \ref{fig:skewed_slip_with_dp} shows a modified situation where the pore throat is inclined by an angle of $\SI{45}{\degree}$ and features an inflow of
$\SI{1.33e-17}{m^3/\second}$ from the bottom of the throat. For this case, the differences between the
coupled models including or not including the slip term are smaller than before. Both deviate from the reference solution at the left half of the image,
close to the interface. The pronounced upward flow of at the left edge of the throat is not captured by the coupled models.
Nevertheless, including the slip term slightly improves the fit with the reference solution. The relative error for the velocity given by Eq. \eqref{eq:norm} reduces by a factor of 1.3 from \SI{1.07e-02}{} to \SI{8.14e-03}{} after including the slip term (compared to a factor of 9 for the orthogonal throat). The error for pressure decreases from \SI{4.78e-03}{} to \SI{3.23e-03}{}
which corresponds to a factor of 1.48 compared to 3.72 for the orthogonal throat.\\

Note that we still use $\beta_\mathrm{throat}$ from the case of the orthogonal throat without inflow from the bottom since the idea of the novel approach
is to provide an approximation of the slip velocity with reasonable accuracy under a minimum of complexity and computational cost, thus evaluating $\beta_\mathrm{throat}$ on a multitude
of different geometries and flow configurations is not intended.

Only for sake of a detailed analysis, we re-evaluated $\beta_\mathrm{throat}$ for the inclined flow setup. This yields a new value of of 28001 (compared to 84550) which, however,
produces very similar results when used in the coupled model which is why the latter are not presented here.
This means that for a given inclined inflow, the choice of $\beta_\mathrm{throat}$ seems to have a low effect compared to the influence
of $ [\mathbf{v}]^{\text{PNM}} \cdot  \mathbf{t}_i$.

\begin{figure}[H]
\centering
\includegraphics[width=0.7\textwidth]{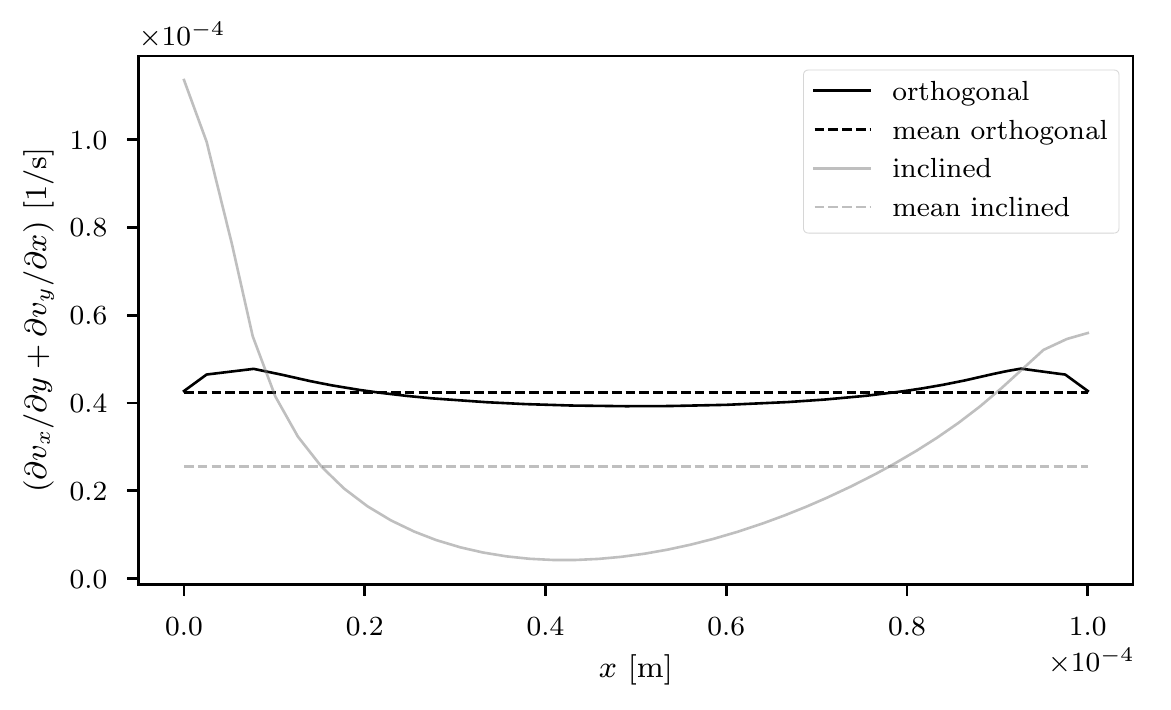}
\caption{Comparison of the shear rates at the interface for the orthogonal setup (Fig. \ref{fig:orthogonal_slip} and the setup featuring an inclined throat with %
         inflow from the bottom (Fig. \ref{fig:skewed_slip_with_dp}).}
\label{fig:slip_gradient}
\end{figure}

Fig. \ref{fig:slip_gradient} shows the shear rate $(\partial v_x / \partial y + \partial v_y / \partial x)$ along the interface for the orthogonal and the inclined throat, as well as the respective averaged values used in Eq. \eqref{eq:beta_throat}. While the curve for the orthogonal throat appears to be entirely
symmetrical with rather small local deviations from the averaged value, the situation is quite different for the inclined throat with bottom inflow.
Here, the highest shear rates occur at the left edge of the throat and the curve is not symmetrical anymore.
The averaged value seems less adequate for describing the shear behavior in an integral way.

We furthermore repeated the evaluation of $\beta_\mathrm{throat}$ under
the consideration of $ [\mathbf{v}]^{\text{PNM}} \cdot  \mathbf{t}$ which, however, is actually slightly higher than the free-flow slip velocity at the interface, thus leading to negative values of $\beta_\mathrm{throat}$ and nonphysical results in the coupled model.

%In order to asses the influence of the superimpose base flow from left to right in the free flow channel, the setup is further modified in Fig. \ref{fig:skewed_slip_without_dp} where the pressure gradient within the free-flow channel is set to zero,
%i.e, $p_\mathrm{in} = p_\mathrm{out}$. In the reference solution,
%the fluid leaves the throat not only towards the right but there is actually also flow in negative $x$-direction at the left edge of the interface. Both coupled models are not
%able to reproduce this, some velocities vectors even point in an opposite directing compared to the reference solution as shown in the magnified
%section of Fig. \ref{fig:skewed_slip_without_dp}. Again the deviations in both magnitude and orientation of the vectors is less if the slip velocity is included.
%
%\begin{figure}[H]
%\centering
%\includegraphics[width=0.7\textwidth]{./skewed_nodp_withinflow.png}
%\caption{Close-up of the interface region (see gray inlay) for an inclined throat with \mbox{$w_\mathrm{throat} = \SI{100}{\micro \meter}$}. %
%         There is an inflow from the bottom of the throat and no pressure gradient (stagnant flow) in the free-flow channel. %
%         The yellow, purple and black velocity  vectors correspond to the reference (quasi-3D) model, the coupled model including the slip %
%         term and the coupled model with a no-slip condition at the interface.}
%\label{fig:skewed_slip_without_dp}
%\end{figure}

In summary, both parameters for the coupled model, the throat conductance factor $k_{ij}$ and the throat slip coefficient $\beta_\mathrm{throat}$ have been determined
numerically. A power-law functional relation between $\beta_\mathrm{throat}$ and $w_\mathrm{throat}$ could be found. For the
rectangular throat with no inclination, the coupled model including the slip velocity at the interface could accurately reproduce the reference solution for
$w_\mathrm{throat} < \SI{400}{\micro \meter}$ which is a clear improvement compared to the original coupled model \citep{weishaupt2019a}, where a no-slip condition at the
interface was assumed for the given geometry. Modifying the setup by considering an inclined throat with an inflow shows the limitations of the proposed approach. Nevertheless, the coupled model with the inclined throat still benefits slightly from accounting for the slip velocities. Considerable deviations only occur directly at the interface and diminish with increasing distance.
The micromodel geometry under consideration in this paper only features orthogonal throats at the interface, therefore the novel approach is suitable for
increasing the coupled model's accuracy in an simple and efficient manner. In future work, it could be generalized
for various types of geometries and flow configurations.

\subsection{Simulation results of the coupled model}

\begin{figure}[H]
\centering
\includegraphics[width=0.8\textwidth]{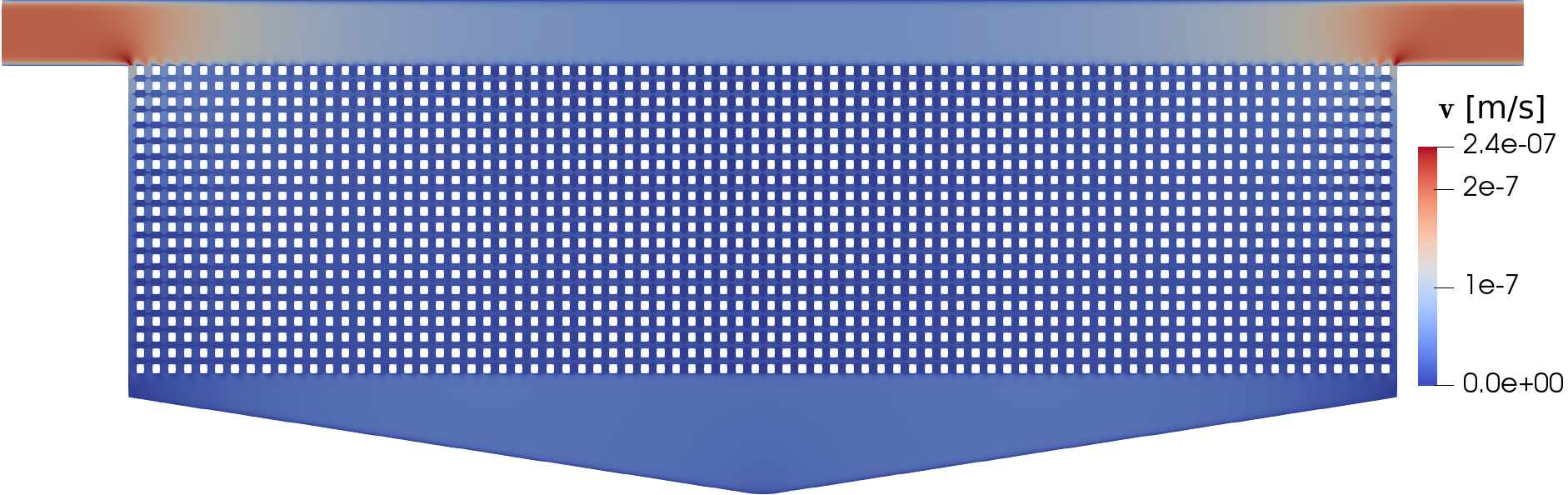} \\ \vspace{4mm}
\includegraphics[width=0.8\textwidth]{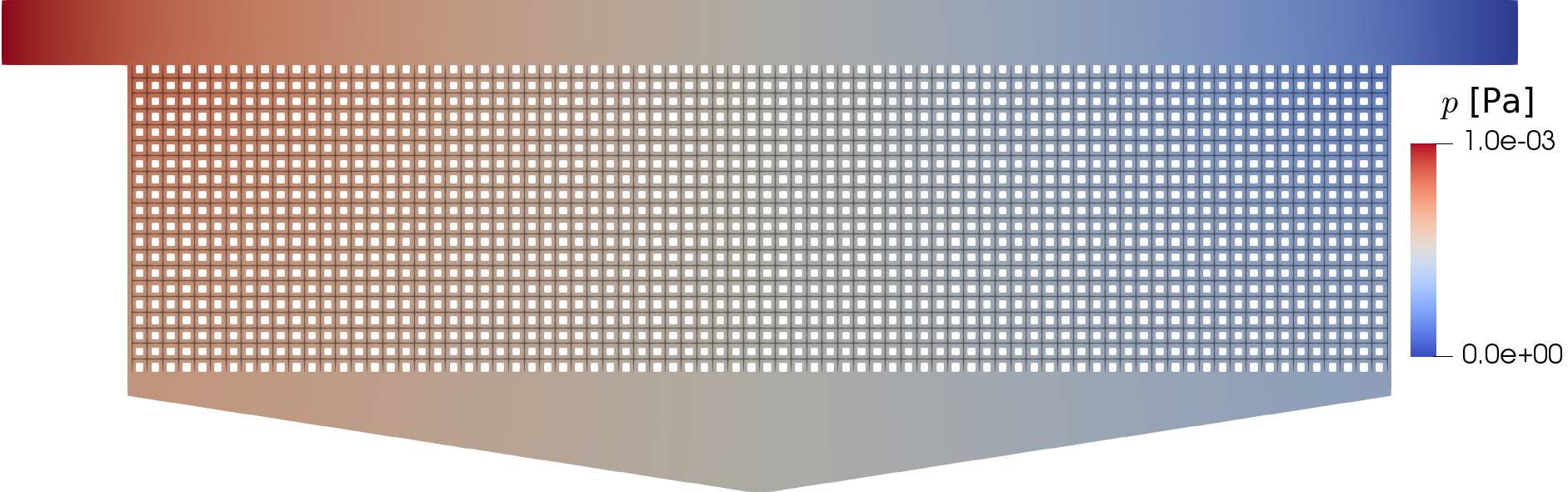}
\caption{Two-dimensional velocity field ($\mathbf{v}_\mathrm{2D}$, top) and pressure field (bottom) obtained by the coupled model corresponding %
         to the center plane ($z= \SI{100e-6}{\meter}$) of the 3D model. The one-dimensional throat elements of the pore-network model have been extruded %
         for visualization purposes using ParaView's \citep{ayachit2015a} \textit{Tube} filter. The pore throats show an averaged velocity based on Eq. \eqref{eq:v_throat}.}
\label{fig:coupled_v_p}
\end{figure}

In this last section, we present the results of the hybrid-dimensional, coupled model. For sake of comparability, the previously shown results of the quasi-3D model
(which is also a component of the coupled model) will serve as reference solutions here. While the free-flow channel and the triangular reservoir are still discretely
meshed and resolved in the same way as before, the porous domain is now substituted by an equivalent pore-network model whose input parameters have been chosen based on
the numerical upscaling procedure described above, i.e., $w_\mathrm{throat} = l_\mathrm{throat} = \SI{240}{\micro \meter}$, \mbox{$k_{ij,t} = \SI{3.05e-10}{m^3/(s \pascal)}$},
\mbox{$k_{1/2, i} = k_{1/2, j} = \SI{8.47e-10}{m^3/(s \pascal)}$} and $\beta_\mathrm{throat} = \SI{33000}{\per \meter}$. Again, the same boundary conditions are chosen as before while all
pore bodies on the boundary, except those at the interfaces, feature Neumann no-flow conditions. Running the coupled model on a single core of the same machine as before took
less than 5 minutes which is only $\SI{43}{\percent}$ of the CPU time used the quasi-3D model. This is caused by the reduction of the number of degrees of freedom from 9412010 to
3737351 due to the use of the pore-network model in the porous domain. Fig. \ref{fig:coupled_v_p} presents the corresponding velocity and pressure fields for the coupled model.
Note that the pore throats show averaged velocities based on Eq. \eqref{eq:v_throat} which is by implication smaller than the peak free-flow velocities at the associated
interface.

\begin{figure}[H]
\centering
\includegraphics[width=1.0\textwidth]{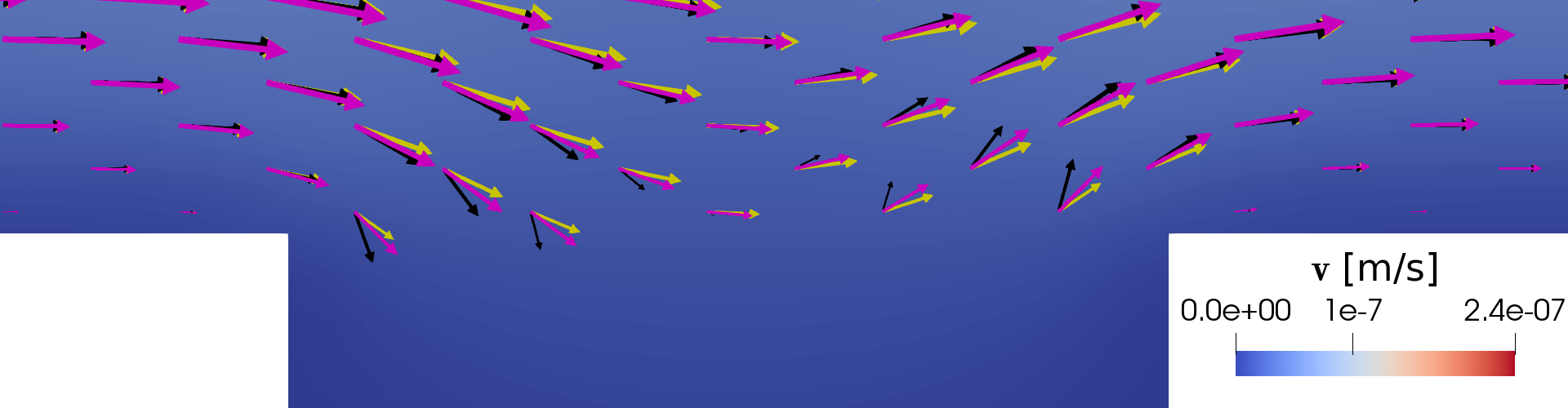}
\caption{Close-up of the interface region at the central throat ($x/l= 80.5, y/l = 0$).
         The yellow, purple and black velocity  vectors correspond to the reference (quasi-3D) model, the coupled model including the slip %
         term and the coupled model with a no-slip condition at the interface.}
\label{fig:coupled_zoom_center}
\end{figure}

In Fig. \ref{fig:coupled_zoom_center}, the central throat intersecting with the interface at $y/l = 0$, $x/l = 80.5$ is magnified. The velocity vectors of the quasi-3D reference
solution are given in yellow, those of the coupled model with slip are shown in purple while those of the coupled model assuming a no-slip condition at the throat openings a marked black.
The main channel flow slightly dips into the throat cavity on the left just to re-enter the main channel on the right. There is no net mass flux across the interface. The flow behavior
is generally reflected by all three models. However, there is a significantly higher agreement between the reference solution's vectors and the one of the coupled model with slip, both in magnitude and
orientation. The vectors' $y$-component of both coupled models is essentially determined by the coupling condition for the conservation of momentum in normal direction,
Eq. \eqref{eq:coupling_normal}, and thus more or less identical. The black vectors, however, which correspond to the coupled model assigning a no-slip condition at the throats, feature
drastically decreased $x$-components.

\begin{figure}[H]
\centering
\includegraphics[width=1.0\textwidth]{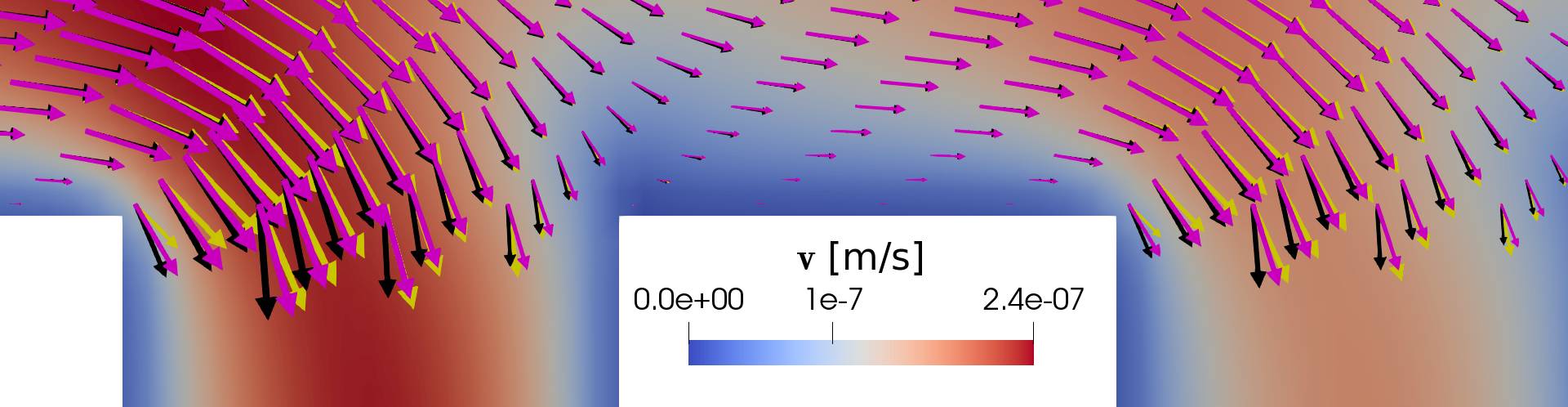}
\caption{Close-up of the interface region at the two leftmost throats ($0 \leq x/l \leq 3, y/l = 0$).
         The yellow, purple and black velocity  vectors correspond to the reference (quasi-3D) model, the coupled model including the slip %
         term and the coupled model with a no-slip condition at the interface.}
\label{fig:coupled_zoom_left}
\end{figure}

The same pattern can be observed in Fig. \ref{fig:coupled_zoom_left} which shows a close-up of the two leftmost throats at the interface. Here we have a pronounced downward flow
from the free-flow channel into the porous domain. Again there is a much better match with the reference solution  if the slip velocity is taken into account.
Using Eq. \ref{eq:norm} to calculate the relative error, based on the velocities in the free flow channel and the triangular region, yields values of
\SI{9.49e-03}{} for the coupled model including the slip term and \SI{1.76e-02}{} for the coupled model without the slip term, both with respect to the quasi-3D model simulation
results as discussed before. In analogy, the relative error values for the pressure are \SI{3.72e-04}{} and \SI{4.52e-04}{}.

\begin{figure}[H]
\centering
\includegraphics[width=1.0\textwidth]{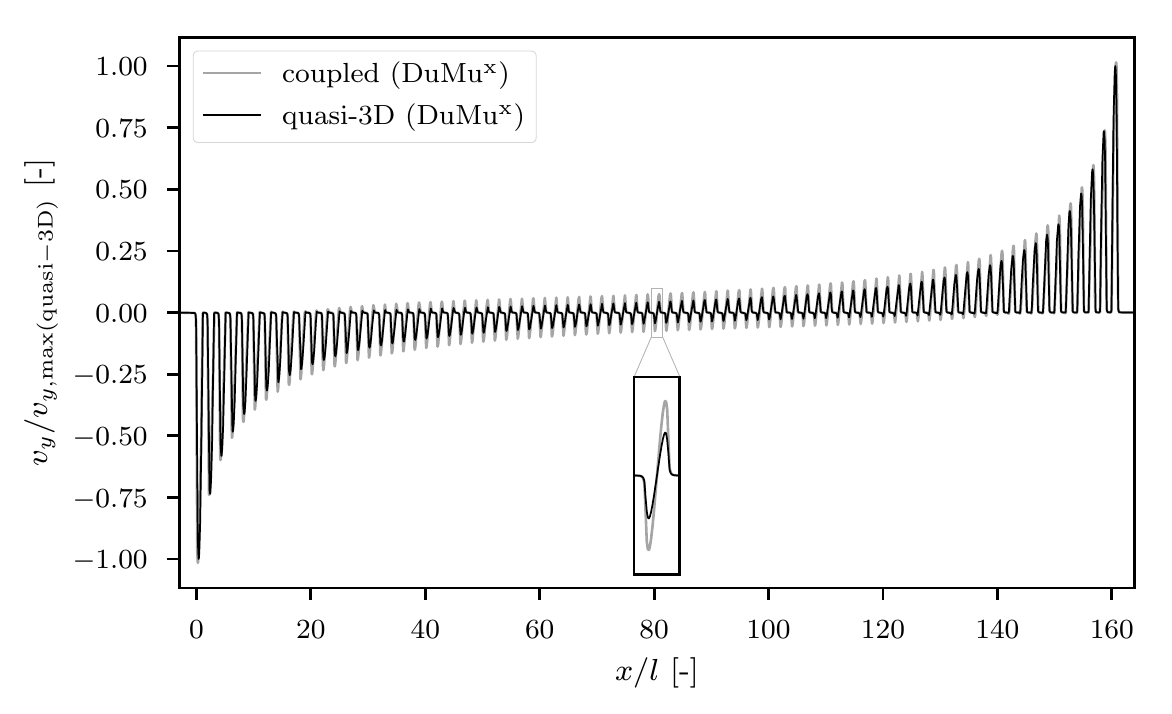}
\caption{Normalized $v_y$ at the interface in stream-wise direction ($y/l=0$) for the quasi-3D and the coupled model.}
\label{fig:comparison_dumux_coupled_interface}
\end{figure}

\begin{figure}[H]
\centering
\includegraphics[width=1.0\textwidth]{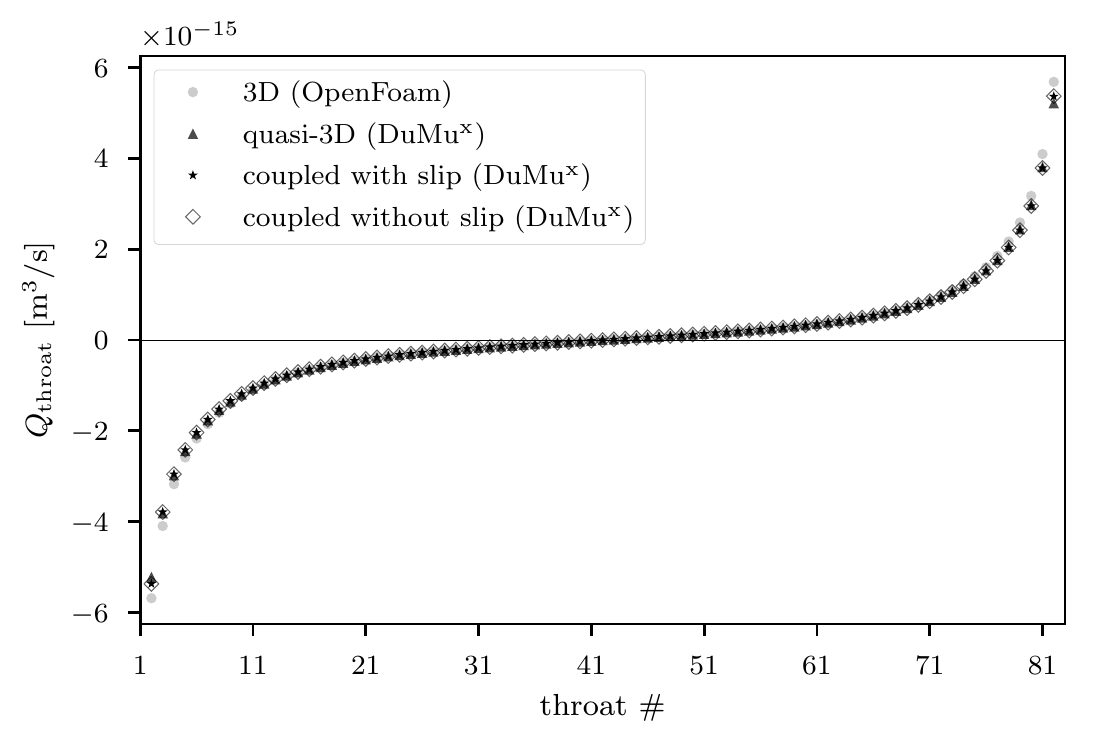}
\caption{Discrete volumetric flow rates at all throats intersecting with the interface for all numerical models.}
\label{fig:throat_fluxes}
\end{figure}

In Fig. \ref{fig:comparison_dumux_coupled_interface}, the vertical velocities $v_y$ over the length of the interface are presented. The solution of the coupled model qualitatively follows the
one of the quasi-3D model, while generally over-predicting the velocity peaks, especially at the center of the model where discrepancies up to $\SI{80}{\percent}$ can be found.
The large mismatches only occur near the edges of the throat openings. This can be explained by the symmetric downwards and upwards flows
as previously described with the help of Fig. \ref{fig:coupled_zoom_center}, which is also reflected in the zoom-in shown of Fig. \ref{fig:comparison_dumux_coupled_interface}.
As the over- and undershoots tend to cancel out at each throat, the global mass transfer over the interface does not suffer noticeably which is underlined by Fig. \ref{fig:throat_fluxes}.

Here, the total volumetric flow through each throat at the interface is shown, as evaluated by Eqs. \eqref{eq:3d_flow_rate} and \eqref{eq:2d_flow_rate}.

 The throats are label from left to right from \#1 to \#81.
The values of the coupled models are almost identical to the ones of the quasi-3D reference solution, regardless whether slip is considered or not which means that the vertical
mass exchange between the free flow and porous medium is not significantly influenced by the slip velocity above the throats.

For comparison, also the throat fluxes of the 3D simulation conducted with OpenFoam are shown in Fig. \ref{fig:throat_fluxes}. The latter are in close accordance with the values for
the quasi-3D and the coupled models implemented in \dumux. The deviations are largest at the very left and very right throats which goes in line with the findings concerning
Fig. \ref{fig:openfoam_analytical_v_y} and Fig. \ref{fig:dumux_v} previously discussed.

\begin{figure}[H]
\centering
\includegraphics[width=0.9\textwidth]{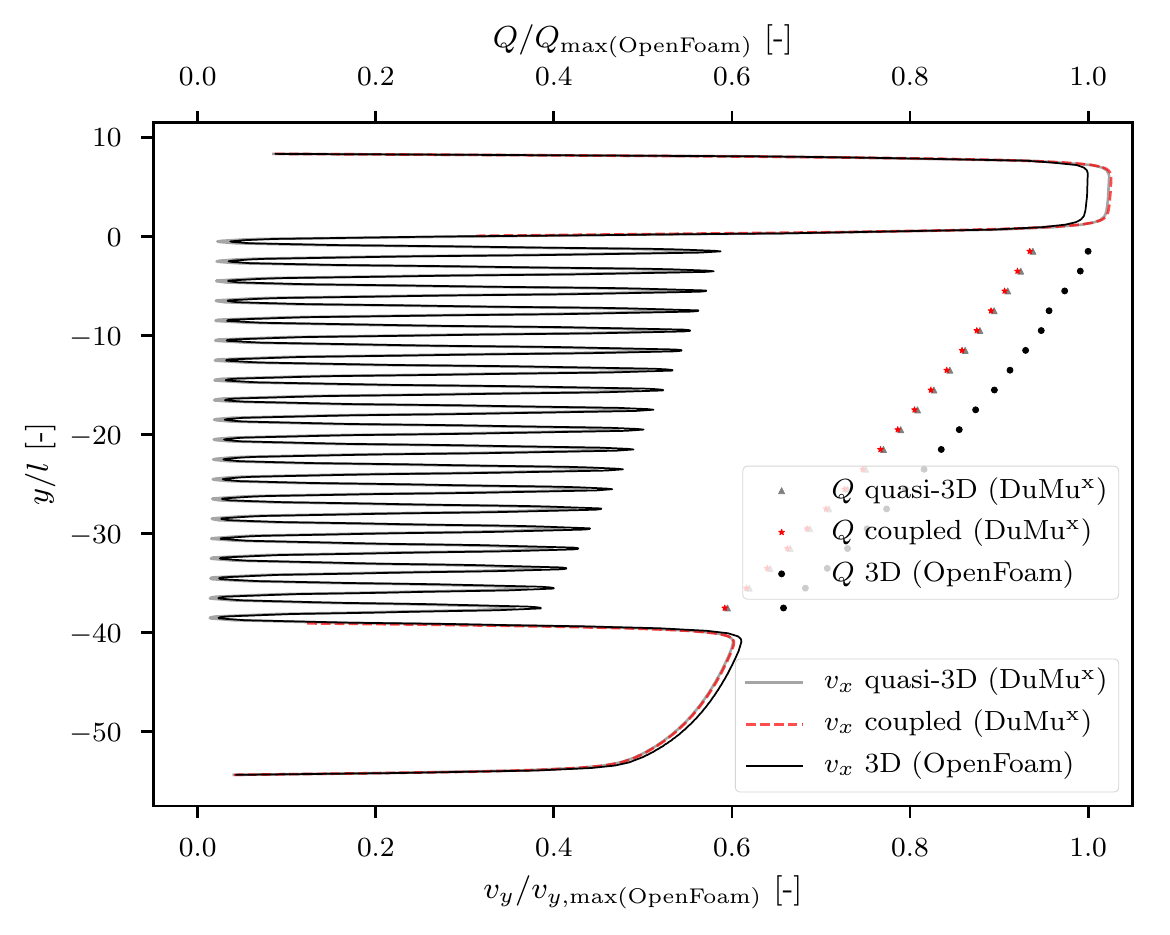}
\caption{Velocity profiles $v_x$ over $y$ at $x/l = 80.5$  and discrete volumetric flow rates at $x/l = 79.5$ for all numerical models,
         normalized by the maximum values of OpenFoam.%
         The coupled model only features continuous velocities in the free-flow channel and the triangular region.}
\label{fig:throat_fluxes_hor}
\end{figure}

Finally, Fig. \ref{fig:throat_fluxes_hor} sheds some light onto the flow conditions within the porous medium at the vertical center line of the
micromodel. Depicted are the normalized horizontal velocities at $x/l = 80.5$, which is the exact center of the porous medium, and the
integral volume fluxes $Q$ at the throats directly left to the center line at $x/l = 79.5$, likewise normalized and evaluated by Eqs. \eqref{eq:3d_flow_rate}, \eqref{eq:2d_flow_rate} and \eqref{eq:pnm-simple}. As the pore-network model only
yields averaged velocities within the pore throats, $v_x$ is only drawn in the free-flow channel and the triangular region, where it matches
almost perfectly the solution of the quasi-3D model. Both models also give rise to very similar integral volume fluxes within the throats, which
deviate by around \SI{6}{\percent} from the values of the 3D simulation. This can be explained by
the aforementioned unfavorable aspect ratio of 0.83 in the pore throats which impairs the accuracy of
Eqs. \eqref{eq:flekkoy} and \eqref{eq:2d_flow_rate} used for the quasi-3D model from which subsequently also the throat conductances
were derived by numerical upscaling, as described in Sec. \ref{sec:upscaling}.

In summary, we have shown that the coupled, hybrid-dimensional model is able to reproduce the reference quasi-3D results both qualitatively and quantitatively. Pronounced local velocity
deviations occur locally at the edges of the throat openings which, however, do not considerably impair the mass exchange across the interface between free flow and porous medium.
Including a mechanism for slip velocities above the throats noticeably improves the results of the coupled model and comes with no additional run-time cost. The coupled model yields a
speedup of 2.3 compared to the quasi-3D model.

\section{Summary and Conclusions}
\label{sec:conclusion}
In this work, we have presented a model reduction approach for simulating free flow over an adjacent porous medium.
A full three-dimensional reference solution for the entire computational domain was created using the open-source CFD toolbox OpenFoam which required over 60 million grid cells and five hours of CPU time on 30 cores. The results of this model have been validated against experimental micro-PIV data \citep{terzis2019a}, showing very good agreement over the entire measured domain.
Extending the numerical model for the inclusion of microscopic geometrical features such as surface roughness \citep{silva2008a} or pillar edge fillets will be addressed in future work.

As a first step of a two-fold model reduction,
a quasi-3D model is implemented in the open-source simulator \dumux, where an additional friction term \citep{flekky1995a} accounts for the omitted dimension in $z$-direction.
A close match with the OpenFoam results is found with local deviations always less than \SI{8}{\percent}. The required CPU time dropped to around 11 minutes on a single core. This is a massive decrease compared to the five hours of CPU time on 30 cores needed by OpenFoam. However, one needs to take into account the different solution strategies followed by the different solvers: while in \dumux, a stationary problem was solved on a single core, OpenFoam's \texttt{icoFoam} solver is inherently transient but parallelizable which makes the CPU times only partially comparable.

Following the reduction by one dimension, a pore-network model is introduced to account for the porous domain, based on earlier work presented in \cite{weishaupt2019a}. Here, the coupling conditions between the two models are refined in the sense that now also
slip velocities at the intersection between the pore throats and the free-flow domain are accounted for. This was not possible in the original implementation where
no-slip coupling conditions would always hold for the given geometric setup. Including this slip mechanism greatly improves the accuracy of the coupled model both locally
and globally at no additional run-time cost. To this end, a geometry-specific material parameter had to be evaluated numerically. A power-law fit was found for a simple approximation of
the required parameter for a wide range of throat widths. The new approach works best for throats that intersect orthogonally with the interface and has some limitations
for inclined throats featuring in- or outflow. Nevertheless, including the slip mechanism still improves the simulation results for inclined throats compared to the previous
implementation. If needed, a more general, but potentially more complex approximation of the slip velocities might be elaborated on in future work, possibly under the
consideration and generalization of existing analytical expressions \citep{moffatt1964a, jeong2001a} and accounting for the orientation and intensity of the
flow across the interface \citep{yang2019a}. However, for the geometry at hand, the presented approach proved
to be simple, efficient and sufficiently accurate. A speedup of 2.3 compared to the quasi-3D model was observed which could be further increased by means of local grid refinement in the
free-flow region \citep{vittoz2017a}. In summary, the coupled, hybrid-dimensional model
is an interesting and efficient option for the simulation of coupled systems of free-flow over an
permeable medium. It can be certainly used as a powerful design tool during the optimization of microfluidic experiments as well as in industrial applications providing accurate results in a timely manner.

\section*{Acknowledgments}
We thank the Deutsche Forschungsgemeinschaft (DFG, German Research Foundation) for supporting this work by funding SFB 1313, Project Number 327154368.
Guang Yang is grateful to the support from the National Natural Science Foundation of China (NSFC).
We would also like to thank Ivan Yotov and Wietse Boon for fruitful discussions.

\bibliography{literature}
\bibliographystyle{elsarticle-num}

\end{document}